\documentclass[aps,pra,superscriptaddress,reprint,a4paper,longbibliography,bibnotes]{revtex4-2}

\usepackage[utf8]{inputenc}
\usepackage[T1]{fontenc}
\usepackage{amsmath,amsfonts,amssymb}
\usepackage{mathtools} 
\usepackage{graphicx}
\usepackage{bm}
\usepackage{xcolor}

\usepackage{braket}

\usepackage{float}

\usepackage{algorithm}
\usepackage{algorithmic}

\usepackage{changes}

\definecolor{myurlcolor}{rgb}{0,0,0.7}
\definecolor{myrefcolor}{rgb}{0.8,0,0}
\usepackage{hyperref}
\hypersetup{
    unicode=true, bookmarksnumbered=false, bookmarksopen=false, breaklinks=false, pdfborder={0 0 0},
    colorlinks=true,
    linkcolor=myrefcolor,citecolor=myurlcolor,urlcolor=myurlcolor
}

\newcommand{\trace}{\mathrm{Tr}}

\usepackage{xparse}
\NewDocumentCommand\Tr{g}{
	\IfNoValueTF{#1}
	{\trace}
	{\trace\!\left\{#1\right\}}
}

\def\dd{\mathrm d}

\def\ee{\mathrm e}

\newcommand{\trm}[1]{\textrm{#1}}
\newcommand{\mrm}[1]{\mathrm{#1}}

\newcommand{\eref}[1]{(\ref{#1})}

\newcommand{\eqnref}[1]{Eq.~(\ref{#1})}

\newcommand{\figref}[1]{Fig.~\ref{#1}}

\newcommand{\secref}[1]{Sec.~\ref{#1}}

\newcommand{\kappad}{{\kappa_{\mrm{d}}}}
\newcommand{\kappal}{{\kappa_{\mrm{l}}}}

\NewDocumentCommand\av{m+g}{%
	\IfNoValueTF{#2}
	{\left\langle{#1}\right\rangle}
	{\left\langle{#1}\right\rangle_{#2}}%
}
\NewDocumentCommand\Var{m+g}{%
	\IfNoValueTF{#2}
	{\mathrm{Var}\!\left[#1\right]}
	{\mathrm{Var}\!\left[#1\right]_{#2}}%
}

\newcommand{\kB}{k_\text{B}}

\makeatletter
\newcommand{\vast}{\bBigg@{4}}
\newcommand{\Vast}{\bBigg@{5}}
\makeatother

\begin{document}

\title{Efficient inference of quantum system parameters by Approximate Bayesian Computation}

\author{Lewis A.~Clark}
\affiliation{Department of Optics, Palacky University, 17. listopadu 1192/12, 77146 Olomouc, Czechia}
\affiliation{Centre for Quantum Optical Technologies, Centre of New Technologies, University of Warsaw, Banacha 2c, 02-097 Warsaw, Poland}
\author{Jan Ko\l{}ody\'nski}
\email{jankolo@ifpan.edu.pl}
\affiliation{Institute of Physics, Polish Academy of Sciences, Aleja Lotnik\'{o}w 32/46, 02-668 Warsaw, Poland}
\affiliation{Centre for Quantum Optical Technologies, Centre of New Technologies, University of Warsaw, Banacha 2c, 02-097 Warsaw, Poland}

\begin{abstract}
	The ability to efficiently infer system parameters is essential in any signal-processing task that requires fast operation. Dealing with quantum systems, a serious challenge arises due to substantial growth of the underlying Hilbert space with the system size. As the statistics of the measurement data observed, i.e.~the \emph{likelihood}, can no longer be easily computed, common approaches such as maximum-likelihood estimators or particle filters become impractical. To address this issue, we propose the use of the Approximate Bayesian Computation (ABC) algorithm, which evades likelihood computation by sampling from a library of measurement data---\emph{a priori} prepared for a given quantum device. We apply ABC to interpret photodetection click-patterns arising when probing in real time a two-level atom and an optomechanical system. For the latter, we consider both linear and non-linear regimes, in order to show how to tailor the ABC algorithm by understanding the quantum measurement statistics. Our work demonstrates that fast parameter inference may be possible no matter the complexity of a quantum device and the measurement scheme involved.
\end{abstract}
\maketitle

\section{Introduction}
%
\emph{Quantum statistical inference}~\cite{BarndorffNielsen2003} forms an essential part of any sensing scenario, in which a quantum system is used to probe external parameters affecting its dynamics~\cite{Giovannetti2004,Paris2009,DemkowiczDobrzanski2015,Yuan2017}. It lies at the core of quantum metrology protocols~\cite{Giovannetti2001,Giovannetti2006,Giovannetti2011}, which may then offer unprecedented sensitivity thanks to genuine quantum effects, e.g.~entanglement~\cite{Pezze2018} or others~\cite{Braun2018}, exhibited by the sensor. This has led to breakthroughs in construction of ultra-precise clocks~\cite{Colombo2022}, magnetometers~\cite{Budker2007}, interferometers~\cite{Cronin2009,Bongs2019}, or even gravitational-wave detectors~\cite{Tse2019}. The latter, in fact, constitute a seminal example in which signal extraction of cosmic events may only be possible by using state-of-the-art sampling-based inference techniques~\cite{Veitch2015,Abbott2020}.

What is more, the gravitational-wave detector~\cite{Yu2020} may be viewed as an optomechanical device~\cite{Aspelmeyer2014} operated in a two-stage transducing architecture, in which an external signal (gravitational wave) affects a well-isolated quantum sensor (mirrors) that is continuously measured (optically) in real time~\cite{wieczorek_optimal_2015,rossi_observing_2019}. The same picture applies to atomic sensors~\cite{Budker2007}, in which it is rather the atomic spin that is sensitive to external magnetic fields, while being simultaneously monitored by the light~\cite{Geremia2003,Moelmer2004,albarelli_ultimate_2017,amoros-binefa_noisy_2021}. The formalism of continuous quantum measurements~\cite{Belavkin1989,Plenio1998,Jacobs2014} allows one then to build a model describing the detection process as well as (conditional) sensor dynamics, based on which statistical inference can be used to interpret the measured data and extract the signal being sensed~\cite{Gammelmark2013,kiilerich_estimation_2014,Kiilerich2016}. Moreover, if this can be done \emph{efficiently}, i.e.~fast, the information may be used `on the fly' to further enhance the process by controlling the device in real time and apply active feedback~\cite{Nurdin2017,Hamerly2012,Yamamoto2014}. This has been spectacularly demonstrated in optomechanical devices~\cite{wilson_measurement-based_2015,rossi_measurement-based_2018,sudhir_appearance_2017,Ernzer2023}, and also ones involving levitated nanoparticles~\cite{Setter2018,magrini_real-time_2021}, when probed by optical homodyning~\cite{Kiilerich2016}. In such a case,
\begin{figure}[H]
	\includegraphics[width=\columnwidth]{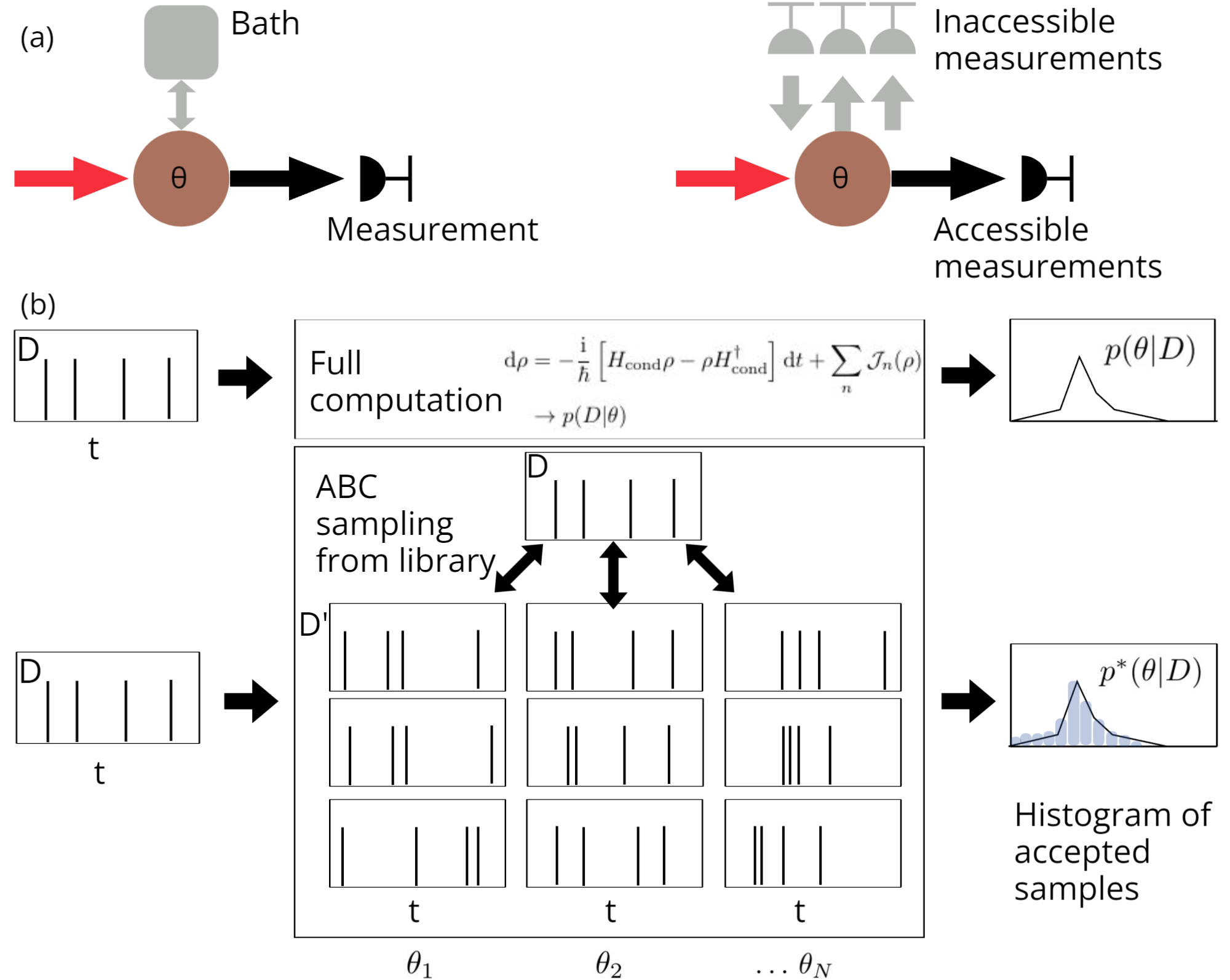}
	\caption{
	\textbf{Parameter inference in the quantum setting.} (a)~A parameter $\theta$ is encoded onto a quantum system during its evolution. 
	To infer the parameter value, the system is probed by an external field, e.g.~light, (red arrow) that is subsequently measured (black arrow). In parallel, the system experiences dissipation and excitation due to interactions with a bath. As long as the impact of the bath on the system is memoryless (Markovian), it can always be modelled as interactions with virtual external fields whose measurements are, however, \emph{inaccessible}.  (b)~The value of $\theta$ is inferred from the \emph{accessible} data $D$---here, a pattern of photon-clicks observed at distinct times---by reconstructing the \emph{posterior distribution} $p(\theta | D)$. This requires determining  explicitly the form of the \emph{likelihood} $p(D|\theta)$ for any $\theta$, given the specific dynamics
	of the system. Such a `full computation' can be avoided, however, by resorting to the \emph{approximate Bayesian computation} (ABC) algorithm that allows one to reconstruct approximately the posterior, $p^*(\theta | D)$, by sampling from a library of detection patterns generated in advance for different values of $\theta$.
	}
	\label{Fig:ABC_setup}
\end{figure}
\noindent as the Gaussianity of the overall dynamical and measurement processes can be assured, statistical inference can indeed be performed quickly, e.g.~by Kalman filtering~\cite{Tsang2009,wieczorek_optimal_2015,JimenezMartinez2018}.

In contrast, the usage of \emph{photon-counting measurements} poses a serious challenge from the statistical inference perspective~\cite{kiilerich_estimation_2014}, which, if overcome, could lead to various breakthroughs with photodetection being recently adopted also for probing optomechanical devices~\cite{cohen_phonon_2015,riedinger_remote_2018,galinskiy_phonon_2020,fiaschi_optomechanical_2021}. 
In \figref{Fig:ABC_setup}(a), we present a scheme in which a general quantum system is probed by light that subsequently undergoes photodetection. Some external or internal parameter, $\theta$, is being encoded onto the system during the probing time, while the system also dissipates due to inevitable interactions with some environment. The goal is to infer the information about $\theta$ from a photon-click pattern, $D$, recorded by detecting emitted photons in real time within some \emph{accessible} channel.

An important tool allowing one to ease the simulation of system dynamics are the quantum Monte Carlo methods~\cite{Dalibard1992,Carmichael1993,Hegerfeldt1993,Molmer1993,Molmer1996}, in which any (Markovian) bath-induced dissipation can be unravelled, i.e.~interpreted as a virtual quantum-jump process yielding emissions into, or excitations from, fictitious \emph{inaccessible} channels, see \figref{Fig:ABC_setup}(a)---although one must (ensemble) average over trajectories of such `unobservable jumps' to obtain the true system dynamics, recording one trajectory is then enough if interested solely in generating a click-pattern of the observed photons, $D$, and ignoring `unobservable clicks'.

Now, as shown in \figref{Fig:ABC_setup}(b), the purpose of Bayesian statistical inference is to reconstruct the \emph{posterior distribution} $p(\theta|D)$~\cite{Trees1968,Kay1993}, which represents most accurately our knowledge about $\theta$ given a particular photon-click pattern $D$ detected. Even if numerically done, this requires, on one hand, sampling from the \emph{likelihood} $p(D|\theta)$ for different values of $\theta$, and on the other, assumes computation of $p(D|\theta)$ to be quick. As the first issue constitutes the main bottleneck in common inference problems with the $\theta$-parameter space being vast, various methods have been designed, such as particle filters~\cite{Murphy2012,Granade2017} or Markov chain Monte Carlo samplers~\cite{Gilks1995,Granade2017} (e.g.~Metropolis-Hastings algorithm~\cite{Gammelmark2013}), to relax it and consider only statistically relevant values of $\theta$. However, within the quantum setting, in which the dimension of the density matrix scales significantly with the system size, already the computation of the likelihood often poses a serious challenge---effectively requiring, see \figref{Fig:ABC_setup}(b), simulations of the system dynamics along a particular trajectory $D$ being observed. Unless the underlying system can be treated as an ensemble of independent constituents of moderate dimension, e.g.~atoms in atomic clocks~\cite{Aeppli2024}, or obeys some symmetry allowing to reduce the effective size of its wavefunction~\cite{Cirac2021,Godley2023}, the likelihood computation becomes impractical, especially for long trajectories with the growth of the integration time, not to mention reaching speeds necessary for applying active inference-based feedback operations~\cite{wilson_measurement-based_2015,rossi_measurement-based_2018,sudhir_appearance_2017,Ernzer2023,Setter2018,magrini_real-time_2021}.

In our work, we demonstrate that the above challenge can be overcome by resorting to so-called \emph{likelihood-free inference methods}, namely \emph{Approximate Bayesian Computation} (ABC)~\cite{Beaumont2002,Turner2012,Sisson2018}, previously applied in the quantum context of Markov processes~\cite{Catana2014} and Hamiltonian learning~\cite{Wiebe2014,Wiebe2014pra}. In particular, as schematically depicted in \figref{Fig:ABC_setup}(b), rather than trying to explicitly reconstruct the posterior $p(\theta|D)$ based on the system dynamical model, one stores \emph{a priori} a sufficiently large library of detection click-patterns generated for different values of $\theta$. Upon observing $D$, the posterior can then be reconstructed by \emph{rejection sampling}, i.e.~by randomly drawing datasets of known parameter $\theta$ from the library, and 
testing them against $D$ based on either key statistical properties (\emph{summary statistics}) that must be appropriately chosen~\cite{Sisson2018}, or by directly comparing them against $D$ with a suitable distance-measure at some extra, but modest, computational cost~\cite{Drovandi2022}.  Despite being approximate, ABC allows one in principle to control the error in reconstructing $p(\theta|D)$~\cite{Sisson2018}---in contrast to methods based on machine learning~\cite{Cimini2021,Khanahmadi2021,Nolan2021,Hsieh2022,Porotti2022,Rinaldi2024}, for example, that are fast once trained, but heuristic~\cite{Murphy2012}.  Although in parameter-estimation tasks it is sufficient to infer only relevant features of the posterior~\cite{Kay1993}, e.g.~dominant moments of the distribution, let us emphasise that our priority here is the verification of ABC and, hence, its accuracy in reconstructing the full posterior.  More precisely, we shall quantify how well the posterior produced by ABC represents its true form, rather than compare its parameter estimation capabilities.

In order to study the capabilities of the ABC algorithm in quantum parameter inference
(posterior reconstruction), we consider three different scenarios:~a two-level atom, a linear optomechanical system, and finally a non-linear optomechanical system; in all of which we consider continuous semi-classical driving and the parameter to be inferred is the detuning of the laser driving from the device. For the two-level atom, the posterior distribution can be derived analytically, and as such there is no benefit of using ABC. Nevertheless, it serves as a useful benchmark for demonstrating the effectiveness of the algorithm and simplicity of its construction. For the linear optomechanical system, only approximate analytic solutions can be obtained, thus to obtain the true posterior numerical methods are required. Hence, the ABC algorithm starts to be useful, although model-based brute-force posterior reconstruction can be considered computable. In stark contrast, operating the optomechanical system in the non-linear system, the computation cost of exhaustive methods becomes unbearable, so that the ABC algorithm becomes essential to reconstruct the posterior distribution in reasonable times. However, the understanding of quantum statistics underlying the measurement data is then crucial for the ABC to be effective.

We organise the paper into the following structure. In Sec.~\ref{Sec:Inference}, we summarise the key concepts of Bayesian inference, before presenting the ABC algorithm and its application. We then demonstrate, in Sec.~\ref{Sec:Trajectories}, how to unravel dissipative dynamics of an open quantum system into respective trajectories while being monitored via photon-counting. Next, in Sec.~\ref{Sec:Models}, we introduce the physical models we consider:~a two-level atom, as well as linear and non-linear optomechanical systems; and apply the theory of Sec.~\ref{Sec:Trajectories} to them. Finally, we study the problem of parameter inference for the three cases and, in particular, the performance of
the ABC algorithm in Sec.~\ref{Sec:ABC}. We conclude our work in Sec.~\ref{Sec:Conclusions}.


\section{Bayesian Inference without likelihood computation} \label{Sec:Inference}
In this section, we overview the fundamentals of Bayesian inference, before highlighting the difficulties in working with it.  We first recall the definition of the Bayesian posterior distribution given some measurement data, in order to then describe how the ABC algorithm operates to reconstruct this approximately, but sufficiently, without the need for explicit computation.

\subsection{Bayesian inference}

As sketched in \figref{Fig:ABC_setup}(b), the purpose of Bayesian inference is to reconstruct the distribution describing our knowledge about a parameter $\theta$ after gathering the measured data $D$. Formally, within Bayesian inference, the unknown parameter is treated as a random variable and its likelihood is calculated over a range of parameter values.  Our initial knowledge about the parameter is included in the form of a prior distribution.  Then, the essential form of Bayesian inference comes from Bayes's rule~\cite{Kay1993}:
\begin{align} \label{Eq:Bayes}
	p(\theta | D) & = \frac{p(D | \theta) p(\theta)}{p(D)}
	= \frac{p(D | \theta) p(\theta)}{\int \dd\theta \, p(D | \theta) p(\theta)} \, ,
\end{align}
which defines the posterior distribution $p(\theta | D)$ based on the likelihood $p(D | \theta)$ and the prior distribution $p(\theta)$.

Importantly, for a given model and measurement process, the likelihood is in principle determinable, however it may be hard to evaluate in practice. Nonetheless, much research in classical estimation theory and statistics has focused on avoiding the computation of the denominator in Eq.~(\ref{Eq:Bayes})~\cite{Murphy2012}. This is because the parameter space of $\theta$ can be extremely large and multi-dimensional, so performing the integral over the whole space constitutes typically the bottleneck in the computation.

\subsubsection{Reconstructing the posterior by sampling}
To mitigate the problem of computing the denominator in Eq.~(\ref{Eq:Bayes}), sampling methods can be used.  The most basic sampling routine is known as {\em rejection sampling}~\cite{Murphy2012}.  Suppose we do not have access to the full posterior $p(\theta | D)$, but are able to efficiently compute some unnormalised version of it, e.g.~$\tilde{p}(\theta | D) = p(D | \theta) p(\theta)$.  While the posterior is automatically normalised with respect to $D$, normalising correctly over the full range of $\theta$ can be computationally challenging without the use of sampling.  This is particularly true for large or multi-dimensional parameter spaces.  To do such sampling, we first choose a (typically simple) {\em proposal distribution} from which the samples of $\theta$ are drawn.  A natural candidate for this in Bayesian inference is the prior $p(\theta)$.  Then, the method is guaranteed to work as long as we may find a constant $M$ (preferably the smallest) such that the inequality $M p(\theta) \geq \tilde{p}(\theta | D)$ is fulfilled for any value of $\theta$. In other words, $M p(\theta)$ provides a valid upper bound on the distribution $\tilde{p}(\theta | D)$, although it may be tight only for some particular values of $\theta$.

The rejection sampling method then proceeds as follows.  We first draw a sample of $\theta$ from the proposal distribution, $p(\theta)$, and then proceed to determine whether it should be accepted or rejected.  In order to do so, we draw some value $u$ from the uniform distribution $U(0,1)$ and accept the sampled $\theta$ if the condition
\begin{align}
	u & \leq \frac{\tilde{p}(\theta | D)}{M p(\theta)}
\end{align}
is satisfied.  As a result, the values of $\theta$ being accepted are actually drawn from the normalised version of the $\tilde{p}(\theta | D)$, i.e.~the true posterior. Hence, by sampling long enough, the full posterior $p(\theta | D)$ can be reconstructed.

The efficiency of this method is determined by the probability of acceptance:
\begin{align}
	p({\rm accept}) = & \int {\rm d} \theta \, \frac{\tilde{p}(\theta | D)}{M p(\theta)} p(\theta) = \frac{1}{M} \int {\rm d} \theta \, \tilde{p} (\theta | D) \, ,
\end{align}
which is highest after choosing the lowest possible $M$ such that the constraint $M p(\theta) \geq \tilde{p}(\theta | D)$ is fulfilled for any $\theta$.

When the parameter space becomes large, rejection sampling can become inefficient, particularly if the proposal distribution (the prior) is not tight to the posterior.  To improve the sampling efficiency, methodologies involving Monte Carlo methods, such as Markov chain Monte Carlo algorithms and particle filters~\cite{Murphy2012,Granade2017} have been developed.  In doing so, the need to search across the whole parameter space can be negated, as the algorithms are designed to focus on areas with higher likelihood.

Nevertheless, the tractability of the problem described here requires that at least the likelihood is simple to obtain.  However, in many complex systems~\cite{Beaumont2002,Catana2014}, such as when the likelihood needs to be computationally generated for a highly variable dataset, this is not the case.  In particular, this applies in the quantum setting within which, unless some analytic shortcuts can be established, computing the likelihood requires full simulation of the dynamics given a particular quantum-jump pattern fixed by the measurement data recorded.  In such a case, it is actually the determination of $p(D|\theta)$ that requires our attention.  To circumvent this, we resort to a branch of statistics known as {\em likelihood free inference}~\cite{Sisson2018}.

\subsection{Approximate Bayesian Computation}
In this work, we focus on a particularly successful technique of likelihood-free inference, known as \emph{Approximate Bayesian Computation} (ABC)~\cite{Sisson2018,Turner2012}.  Before going into the technical details on how the algorithm works, we first motivate the problem as follows.  Suppose we are dealing with a system for which the likelihood of obtaining a particular measurement data is difficult to compute, as it requires to reproduce the (conditional) system dynamics. However, simulating instances of measurement data is relatively easy. In such a case, it should still be possible to efficiently perform inference by sampling, as above, by using a precomputed storage of the measurement data.

In particular, suppose we have a precomputed library of datasets, each generated for a known value of $\theta$.  Then, ABC can be viewed as a variation of the rejection sampling method described above.  Similarly, we start by sampling a value of $\theta$ from the proposal distribution, $p(\theta)$.  Based on this, we choose a subset of the library (dataset pool) containing datasets generated with the sampled $\theta$.  We then draw a dataset $D'$ from this subset and compare it against the dataset observed $D$.  If the datasets are equivalent, we accept the sampled value of $\theta$, i.e.
\begin{align}
	\mathrm{Accept\,if:} \, \, D & = D'\, .
	\label{eq:ABC_accept_ideal}
\end{align}

With sufficient sampling, it is clear to see how this leads to the generation of the posterior. The chance of the dataset being equivalent to the one drawn is proportional to the likelihood of the distribution being generated with that particular value of the parameter.  In the ideal case of each dataset pool (for each value of $\theta$) containing all possible datasets, when sampling over the entire library the probability of acceptance is exactly the likelihood, $p(D|\theta)$,~\cite{Sisson2018}.  Moreover, by letting the proposal distribution be the prior, $p(\theta)$, as in rejection sampling, $\theta$ is effectively drawn from a distribution proportional to the product of these two quantities, i.e.~the posterior.

Clearly though, this procedure is only theoretical, as the probability of the exact dataset to exist in the library is negligibly small. We therefore need to relax the acceptance criteria to be able to obtain solution samples in reasonable times.  This is done in the following two ways.

\subsubsection{Summary statistics}
Instead of comparing the datasets directly, we can instead compare them based on simpler statistics of the distribution.  This for example could be the mean, median, or any moment of the data.  These so-considered \emph{summary statistics} will generally be less informative than the full posterior of the data, but can be efficiently computed for any given dataset $D$.  Furthermore, the summary statistic is said to be \emph{sufficient} for a given dataset $D$, if the posterior distribution can be expressed as a function of the summary statistic rather than the data, i.e.~\cite{Beaumont2010,Turner2012}:
\begin{align} \label{Eq:Sufficient}
	p(\theta | D) =  p(\theta | S(D)) \, .
\end{align}
Although this may not be generally satisfied, a given summary statistic can still provide enough information to distinguish the datesets, so that a reasonable approximation of the true posterior can be provided. Thus, comparing the summary statistics of the dataset to those drawn from the library should be enough if the summary statistics are sufficient, or at least close to being sufficient.  The acceptance rule \eref{eq:ABC_accept_ideal} is then relaxed to
\begin{align}
	\mathrm{Accept\,if:} \, \, S(D) & = S(D')\,,
	\label{eq:ABC_accept_S}
\end{align}
where, for generality, we now let the function $S(\cdot)$ contain multiple statistics, e.g.~higher and higher moments---the more statistics are included, as long as they provide useful information to distinguish between datasets, the closer $S(D)$ is to satisfy the sufficiency condition \eqref{Eq:Sufficient}.

\subsubsection{Relaxation of acceptance criteria}
Requiring that the datasets, or their summary statistics, match exactly is a strict criteria of acceptance that is unlikely to be fulfilled in most cases, especially for large and variable datasets.  The sampling procedure is therefore still inefficient.  To allow for computation in reasonable timescales, we thus relax the acceptance criteria such that we require only that the datasets are close to one another.  More precisely, given some statistical distance $\delta$, the acceptance rule \eref{eq:ABC_accept_S} is further relaxed as follows:
\begin{align} \label{Eq:Accept_ABC}
	\mathrm{Accept\,if:} \, \, \delta\left(S(D),S(D')\right) \leq \epsilon \, ,
\end{align}
where $\epsilon$ is a \emph{threshold of acceptance} that we can choose according to our needs.  For summary statistics with scalar outcomes, we can use the absolute difference as our distance, meaning we have
\begin{align} \label{Eq:Accept_ABC2}
	\mathrm{Accept\,if:} \, \, \left| S(D) - S(D') \right| \leq \epsilon \, .
\end{align}
If multiple statistics are used, the acceptance rule can be defined individually for each, or as some combined statistical distance.  Within this work, we shall use an individual rule for each summary statistic.  In particular, dealing with photon-click patterns $D$ in what follows, we use either the acceptance rule \eqref{Eq:Accept_ABC2} when $S$ is taken to be the \emph{total time} of registering all the clicks, or the rule \eqref{Eq:Accept_ABC} for $S$ chosen to be the \emph{histogram} of inter-click intervals and $\delta$ denoting then the L2-norm.

Generally, the challenge in performing ABC can be found in choosing the summary statistics and then subsequently setting an appropriate threshold.  It is not necessarily obvious what summary statistics are sufficient and moreover, how tight the threshold should be to allow sampling in efficient times.  As such, calibration may be necessary before the algorithm can be implemented on real data. Nonetheless, let us emphasise again that, once performed, the posterior can then be generated fast without need of computing the likelihood, no matter the complexity of the model, as long as a dense library of detection datasets can be \emph{a priori} prepared---recall \figref{Fig:ABC_setup}(b).

The ABC algorithm is summarised below. Given a dataset $D$ representing the observed measurement data, we wish to know the value of the parameter $\theta$ it was generated with. We begin by loading a sufficiently large library of datasets $\mathcal{D}'$, each of which is labelled by its corresponding value of $\theta$ within the range $\Theta$ we wish to consider.  Then, we repeat $\nu$ times the procedure of sampling without replacement a dataset, $D'(\theta)$, from the library $\mathcal{D}'$.  If the summary statistics of the sampled dataset, $S(D'(\theta))$, are within $\epsilon$ of those of the observed one, $S(D)$, as in Eq.~(\ref{Eq:Accept_ABC}), we accept the sample and account for its contribution to the reconstructed posterior distribution. In particular, we add an element to the bin associated with the parameter value $\theta$ labelling $D'(\theta)$. Finally, we normalise the distribution according to the total number of accepted (out of all $\nu$) samples and the width of bins we assumed in coarse-graining the parameter values.
	\begin{algorithm}[H]
		\caption{Summary of ABC}
		\begin{algorithmic}
			\STATE{ {\bf input} $D$, \;$\mathcal{D}'\!=\,\text{Library}[D'(\theta),\theta\in\Theta]$}
			\FOR{$i$ in $\nu$}
				\STATE{Select random $D'(\theta)$ from the library.}
				\IF{$\delta(S(D'(\theta)), S(D)) \le \epsilon$}
					\STATE{$\mathrm{bin}_\theta = \mathrm{bin}_\theta + 1$}
				\ENDIF
			\ENDFOR
			\STATE{Normalise(bin)}
		\end{algorithmic}
		\label{alg:ABC}
	\end{algorithm}

\section{Unravelling dissipative dynamics of continuously monitored quantum systems} 
\label{Sec:Trajectories}
We consider quantum systems whose dissipative dynamics can be described by a master equation of the so-called Gorini-Kossakowski-Sudarshan-Lindblad form, typically referred to as being Markovian~\cite{Breuer2002}, i.e.: 
\begin{align} \label{Eq:Master_general}
	\dot{\rho} = & -\frac{\mathrm{i}}{\hbar} \left[H, \rho \right] + \sum\limits_{n} \Gamma_{n} \left(L_n \rho L_n^\dagger - \frac{1}{2} \left[L_n^\dagger L_n , \rho \right]_+ \right) \, ,
\end{align}
where $\rho \in \mathcal{B}(\mathcal{H}_d)$ is then the density matrix representing the ($d$-dimensional) system at any given time. The Hamiltonian $H$ describes the internal evolution of the system, while the interaction of the system with the bath is described by the Lindblad operators $\left\{L_n\right\}$, with rates $\Gamma_{n}$.  This equation models the \emph{ensemble-average} evolution of a quantum system, i.e.~not distinguishing individual trajectories but rather averaging over them.  If we have a specific measurement record, as in Fig.~\ref{Fig:ABC_setup}(a), we must adapt it to include stochastic elements.

In this paper, we shall consider jump-like processes.  As such, a natural way to re-write Eq.~(\ref{Eq:Master_general}) is
\begin{align}
	\dot{\rho} = & -\frac{\mathrm{i}}{\hbar} \left[H_{\mathrm{cond}} \rho - \rho H_\mathrm{cond}^\dagger\right] + \sum\limits_n \Gamma_n L_n \rho L_n^\dagger \, ,
\end{align}
where $H_{\mathrm{cond}} = H - \sum\limits_n (\mathrm{i} \hbar / 2) \Gamma_n L_n^\dagger L_n$ is a `conditional' non-Hermitian Hamiltonian that describes the effect of the Hamiltonian and processes not leading to excitation in the bath.  The remaining terms correspond to interactions with the bath.  In particular, we point out that the evolution of a system conditioned upon no leakage into the bath can be described by
\begin{align} \label{Eq:No_jump}
	\dot{\rho}_0 = & -\frac{\mathrm{i}}{\hbar} \left[H_{\mathrm{cond}} \rho - \rho H_\mathrm{cond}^\dagger\right] \, ,
\end{align}
with $\rho_0$ being then the unnormalised density matrix, whose trace is the probability of such an evolution, in what is known as the quantum jump approach~\cite{Dalibard1992,Carmichael1993,Hegerfeldt1993,Molmer1993,Molmer1996}.  Motivated by this, we take this {\em unravelling} and write a stochastic master equation, where we distinguish between \emph{accessible} decays (e.g.~photons counted by a detector) and \emph{inaccessible} decays (e.g.~photons missed by a detector, processes that cannot be detected), leading to
\begin{align}
	\dd \rho & = - \frac{\mathrm{i}}{\hbar} \left[H_\mathrm{cond} \rho - \rho H_\mathrm{cond}^\dagger \right] \dd t + \sum\limits_{n \in \mathrm{accessible}} \mathcal{J}_n(\rho) \nonumber \\
	& + \sum\limits_{n \notin \mathrm{accessible}} \Gamma_n L_n \rho L_n^\dagger z \, \dd t \, ,
\end{align}
with the jump-process superoperator:
\begin{align}
	\mathcal{J}_n(\rho) = & \Gamma_n \Tr{L_n^\dagger L_n \rho} \rho \, \dd t + \left( \frac{L_n \rho L_n^\dagger}{\trace\left(L_n^\dagger L_n \rho\right)} - \rho \right) \dd N_t^{n} \,
\end{align}
where $\dd N_t^n$ are increments of a Poisson process, having physical interpretation of a random variable that represents counts over the infinitesimal time $\dd t$ of an emission, with expectation value~\cite{Jacobs2014}:
\begin{align} \label{Eq:Full_weights}
	\left< \dd N_t^{(n)} \right> & = \Gamma_n \, \Tr{L_n^\dagger L_n \rho} \dd t \, .
\end{align}
The size of each expectation value (labelled by $n$) gives then the respective weight of a jump from the Lindblad operator $L_n$.  We call such an unravelling the {\em observed unravelling}.  Note that averaging over the stochastic elements restores the ensemble average provided by the standard master equation in Eq.~(\ref{Eq:Master_general}).

With this in mind, we can equivalently consider an unravelling of all decay channels, both accessible and inaccessible, as pictured in the right of Fig.~\ref{Fig:ABC_setup}(a).  In such a case, we obtain a stochastic element for all channels, leading to the {\em full unravelling}:
\begin{align}
	\dd \rho & = - \frac{\mathrm{i}}{\hbar} \left[H_\mathrm{cond} \rho - \rho H_\mathrm{cond}^\dagger \right] \dd t + \sum\limits_{n} \mathcal{J}_n(\rho) \, ,
\end{align}
which, despite including channels that are not necessarily physical, provides a useful tool in modelling the dynamics of the system.  Following the quantum jump approach~\cite{Dalibard1992,Carmichael1993,Hegerfeldt1993}, we only need to simulate the no-jump part of the dynamics in order to track a trajectory, while applying the stochastic `jump' terms whenever they occur.  Taking into account inaccessible channels, the evolution is a modification of Eq.~(\ref{Eq:No_jump}), i.e.:
\begin{align} \label{Eq:No_accessible_jump}
	\dot{\rho}_{0,\mathrm{acc}} = & -\frac{\mathrm{i}}{\hbar} \left[H_{\mathrm{cond}} \rho - \rho H_\mathrm{cond}^\dagger\right] + \sum\limits_{n \notin \mathrm{accessible}} \Gamma_n L_n \rho L_n^\dagger \, ,
\end{align}
where $\rho_{0,\mathrm{acc}}$ is the unnormalised density matrix conditioned on no accessible jumps.  The evolution can be described entirely by Eq.~(\ref{Eq:No_accessible_jump}).

Solving such an equation is numerically challenging for most quantum systems though, due to the size of density matrix needed for complex systems.  Instead, we could work with Eq.~(\ref{Eq:No_jump}), where the system remains in a pure state at all times, so long as we then subsequently average over the inaccessible channels.  This represents a computational advantage, as only a $d$-dimensional state vector needs to be simulated, rather than a $d\times d$ density matrix.  For large $d$, this represents an important speed-up.  When using Monte Carlo methods~\cite{Molmer1993,Molmer1996}, if we retain only the information about the observed channels, and hence therefore average over the unobserved channels, we restore the same results as using the ensemble evolution over inaccessible bath interactions.

\section{Systems considered and their quantum trajectories}
\label{Sec:Models}
In this section we will introduce the physical models of the systems we shall consider for inference.  In particular, we shall show how quantum trajectories that are based upon these systems and represent the measured photon-click patterns can be generated an analysed, using the theory of the previous section.

\subsection{Two-level atom} \label{Sec:Atom}
We first consider a simple system that we can analyse analytically straightforwardly.  The advantage of this is that as analytical results can be easily attained, thus the performance of the ABC algorithm can be tested efficiently for a range of values when we come to this analysis later.  In particular, we consider a two-level atom subject to semi-classical driving with Rabi frequency $\Omega$ and detuning $\Delta$, whose Hamiltonian then reads~\cite{Breuer2002}:
\begin{align}
	H^\mathrm{atom} = & - \hbar \Delta \sigma^+ \sigma^- + \frac{\hbar \Omega}{2} \left(\sigma^- + \sigma^+ \right) \, ,
	\label{eq:H_atom}
\end{align}
where $\sigma^+$/$\sigma^-$ is the atomic raising/lowering operator.  When coupled to a free radiation field under Born-Markov and rotating-wave approximations~\cite{Breuer2002}, the atom evolves according to the master equation:
\begin{align}
	\dot{\rho} = & - \frac{\mathrm{i}}{\hbar} \left[H^\mathrm{atom} , \rho \right] + \Gamma \left( \sigma^- \rho \sigma^+ - \frac{1}{2} \left[\sigma^+ \sigma^- , \rho \right]_+ \right) \, ,
\end{align}
which is of the (Markovian) form \eqref{Eq:Master_general}. For this system, the conditional Hamiltonian takes the form
\begin{align}
	H^\mathrm{atom}_\mathrm{cond} = & H^\mathrm{atom} - \frac{\mathrm{i} \hbar \Gamma}{2} \sigma^+ \sigma^- \, ,
\end{align}
with the conditional (no-jump) evolution given by
\begin{align} \label{Eq:Cond_evo_atom}
	\dot{\rho}_0 = & - \frac{\mathrm{i}}{\hbar} \left[H_\mathrm{cond}^\mathrm{atom} \rho - \rho {H_\mathrm{cond}^\mathrm{atom}}^\dagger \right] \, .
\end{align}
For this simple system, we consider perfect photon detection and as such there are no unobserved jumps, thus no inaccessible decay channels, meaning the stochastic master equation is
\begin{align} \label{Eq:Stochastic_atom}
	\dd \rho = & -\frac{\mathrm{i}}{\hbar} \left[H_\mathrm{cond}^\mathrm{atom} \rho - \rho {H_\mathrm{cond}^\mathrm{atom}}^\dagger \right] \dd t + \mathcal{J}_\mathrm{atom}(\rho) \, ,
\end{align}
with
\begin{align}
	\mathcal{J}_\mathrm{atom}(\rho) = & \Gamma \, \Tr{\sigma^+ \sigma^- \rho} \rho \, \dd t + \left( \frac{\sigma^- \rho \sigma^+}{\trace\left(\sigma^+ \sigma^- \rho\right)} - \rho \right) \dd N_t \,
\end{align}
and weight
\begin{align} \label{Eq:Atom_weight}
	\left< \dd N_t \right> & = \Gamma \, \Tr{\sigma^+ \sigma^- \rho} \dd t \, .
\end{align}

\begin{figure*}[t]
	\centering
	\includegraphics[width=0.98\linewidth]{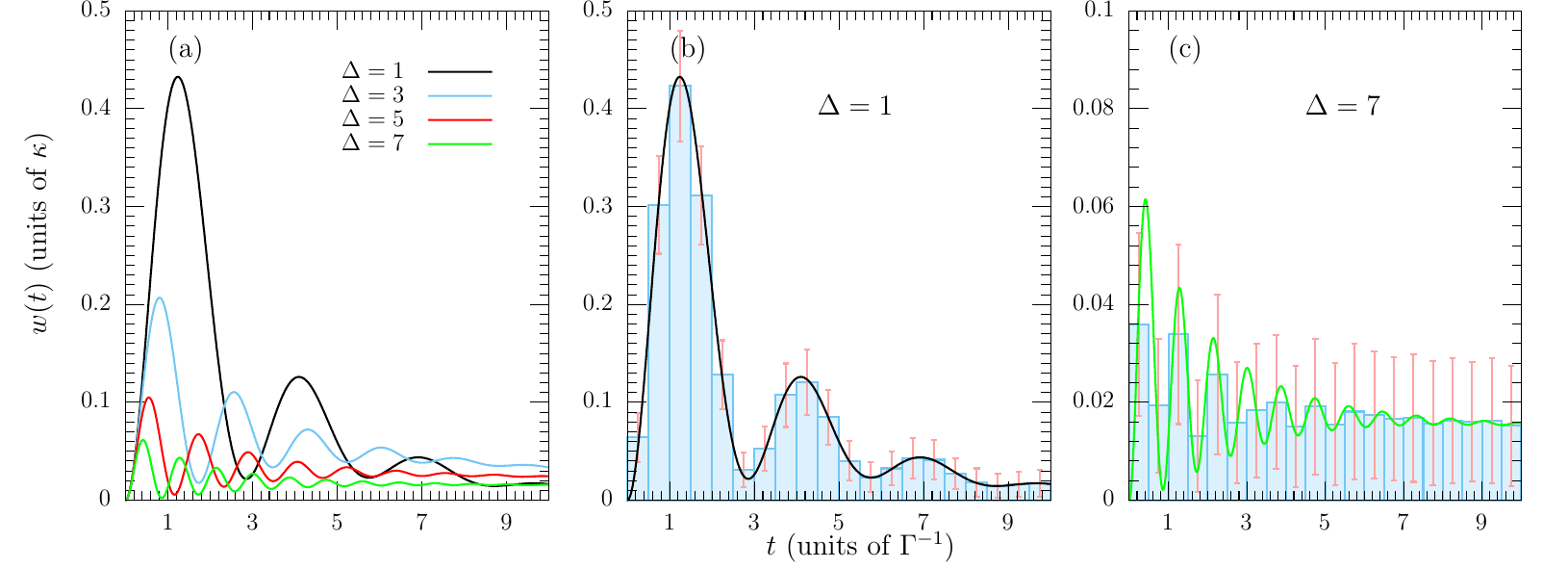}
	\caption{\textbf{Waiting-time distribution for a two-level atom,} whose analytic form is presented in (a) for a number of different values of the detuning $\Delta$.  In (b) and (c) we show the waiting-time distribution compared to the normalised time-binned photon counts averaged over numerically generated trajectories for $\Delta = 1$ and $\Delta = 7$, respectively.  In both (b) and (c) we see strong agreement with the numerics and analytical results.  However, the variance among the trajectories is higher in (c), as marked by the error bars.  This indicates that larger number of samples are required to reconstruct $w(t)$, and naively suggests reduced efficiency of sampling-based methods such as ABC.  For each value of $\Delta$ (expressed in units of $\Gamma$), we take $\Omega = 2 \Gamma$ in \eqnref{eq:H_atom}.}
	\label{Fig:Waiting_atom_analytic}
\end{figure*}

A two-level atom has the useful property that upon emission it always returns to its ground state. As a result, its emission profile constitutes an example of a \emph{renewal process} that is then fully characterised by the \emph{waiting-time distribution}~\cite{Grimmett2001}, i.e.:
\begin{align} \label{Eq:Waiting_atom}
	w(\tau) \dd \tau \, = & P(\text{no click in }[0,\tau]) \nonumber \\
	& \times P(\text{click in }[\tau, \tau + \dd\tau] | \text{no click in } [0 , \tau]) \, .
\end{align}
Substituting in for the two probabilities above, we may write this as~\cite{kiilerich_estimation_2014}:
\begin{align} \label{Eq:Waiting_atom2}
	w(\tau) \dd \tau = & \mathrm{Tr}\left[\rho_0(\tau)\right] \frac{\Gamma \rho^{11}_0(\tau) \dd \tau}{\mathrm{Tr}\left[\rho_0(\tau) \right]} = \Gamma \rho^{11}_0 (\tau) \dd \tau \, .
\end{align}
Following Ref.~\cite{kiilerich_estimation_2014}, but allowing for non-zero detuning $\Delta$, we solve Eq.~(\ref{Eq:Cond_evo_atom}) and obtain the solution for $\rho^{11}_0(t)$.  Hence, using Eq.~(\ref{Eq:Waiting_atom2}) we may directly read off the analytic expression for the waiting-time distribution as \cite{Rinaldi2024}: 
\begin{equation}
	w(\tau)=\frac{\Omega^2}{\Gamma \sqrt{\Xi}} \ee^{-\frac{\Gamma \tau}{2}} 
	\left[\cosh\!\left(\frac{\Gamma \tau}{2} \sqrt{\zeta_+}\right)-\cos\!\left(\frac{\Gamma \tau}{2} \sqrt{\zeta_-}\right)\right],
	\label{eq:wait_time_two-level}
\end{equation}
after defining $\Xi\coloneqq\frac{1}{4}+2\chi_-+4\chi_+^2$ and $\zeta_\pm\coloneqq\sqrt{\Xi}\pm(\frac{1}{2}-2\chi_+)$ with $\chi_\pm\coloneqq(\Delta^2\pm\Omega ^2)/\Gamma ^2$. In Fig.~\ref{Fig:Waiting_atom_analytic}(a), we plot the resulting waiting-time distribution \eref{eq:wait_time_two-level} as function of time $t$ for various detunings $\Delta$.

\subsection{Optomechanical system} \label{Sec:Opto_equations}
The second physical model we consider is an optomechanical system.  The relevant Hamiltonian for an optomechanical system is~\cite{Aspelmeyer2014}:
\begin{align} \label{Eq:H_OM}
	H^\mathrm{OM} = & - \hbar \Delta a^\dagger a + \hbar \omega_\mathrm{M} b^\dagger b + \frac{\hbar \Omega}{2} \left(a + a^\dagger \right) \nonumber \\
	& + \hbar g a^\dagger a \left(b + b^\dagger \right) \, ,
\end{align}
where $a(b)$ is the bosonic annihilation operator for the photonic (phononic) mode of the system, $\Delta$ is the laser detuning, $\omega_\mathrm{M}$ is the mechanical frequency, $\Omega$ is the laser Rabi frequency and $g$ is the optomechanical coupling strength.

Unlike the atom, we now consider two interacting quantum systems with multiple decay channels.  The optical part of the system decays through photon emission, while the mechanical part interacts with a thermal bath, leading to both phonon emission and absorption.  Following the procedure in Sec.~\ref{Sec:Trajectories}, we write a (thermal) master equation as~\cite{Breuer2002}:
\begin{align}
	\dot{\rho} = & -\frac{\mathrm{i}}{\hbar} \left[ H^\mathrm{OM} , \rho \right] + \kappa \left(a \rho a^\dagger - \frac{1}{2} \left[ a^\dagger a , \rho \right]_+ \right) \nonumber \\
	& + \gamma \left(\bar{m} + 1 \right) \left(b \rho b^\dagger - \frac{1}{2} \left[b^\dagger b , \rho \right]_+ \right) \nonumber \\
	& + \gamma \bar{m} \, \left(b^\dagger \rho b - \frac{1}{2} \left[b b^\dagger , \rho \right]_+ \right) \, .
	\label{eq:master_optomech}
\end{align}
We label the optical decay rate $\kappa$ and mechanical decay rate $\gamma$, while we assume the thermal bath interacting with the mechanics has mean occupation $\bar{m}=[\exp(\hbar\omega_\mathrm{M}/\kB T)-1]^{-1}$~\cite{Breuer2002,Clark2022}.  We next write a conditional Hamiltonian
\begin{align}
	H^\mathrm{OM}_{\mathrm{cond}} = & H^\mathrm{OM} - \frac{\mathrm{i} \hbar}{2} \left(\kappa a^\dagger a + \gamma \left(\bar{m} + 1\right) b^\dagger b + \gamma \bar{m} \, b b^\dagger \right) \, .
\end{align}
We now only have one observed decay channel, through finite-efficiency photon counting.  This leads us to an observed unravelling of
\begin{align} \label{Eq:Opto_stochastic_observed}
	\dd \rho = & - \frac{\mathrm{i}}{\hbar} \left[H_\mathrm{cond}^{\mathrm{OM}} \rho - \rho {H_\mathrm{cond}^{\mathrm{OM}}}^\dagger \right] \dd t + \mathcal{J}_\kappad (\rho) \nonumber \\
	& + \left(\kappal a \rho a^\dagger + \gamma \left(\bar{m} + 1 \right) b \rho b^\dagger + \gamma \bar{m} \, b^\dagger \rho b\right) \dd t \, ,
\end{align}
with a jump term:
\begin{align}
	\mathcal{J}_\kappad (\rho) & = \kappad \Tr{a^\dagger a \rho} \rho \, \dd t + \left( \frac{a \rho a^\dagger}{\trace\left(a^\dagger a \rho\right)} - \rho \right) \dd N_t^{\kappad} \, , \nonumber \\
	\left< \dd N_t^\kappad \right> & = \kappad \, \Tr{a^\dagger a \rho} \dd t \, .
\end{align}
In these equations, we have introduced separate decay rates for the detected ($\kappad$) and lost ($\kappal$) photons, such that $\kappad + \kappal = \kappa$.~\cite{Clark2022}.  Finally, we can write the full unravelling as
\begin{align} \label{Eq:Opto_stochastic_full}
	\dd \rho = & -\frac{\mathrm{i}}{\hbar} \left[H_\mathrm{cond}^\mathrm{OM} \rho - \rho {H_\mathrm{cond}^\mathrm{OM}}^\dagger \right] \dd t \nonumber \\
	& + \mathcal{J}_\kappad (\rho) + \mathcal{J}_{\kappal} (\rho) + \mathcal{J}_{\gamma_-} (\rho) + \mathcal{J}_{\gamma_+} (\rho) \, ,
\end{align}
where
\begin{align}
	\mathcal{J}_{\kappal} (\rho) = & \kappal \Tr{a^\dagger a \rho} \rho \, \dd t + \left( \frac{a \rho a^\dagger}{\trace\left(a^\dagger a \rho\right)} - \rho \right) \dd N_t^{\kappal} \, , \nonumber \\
	\mathcal{J}_{\gamma_-} (\rho) = & \gamma \left(\bar{m} + 1 \right) \Tr{b^\dagger b \rho} \rho \, \dd t + \left( \frac{b \rho b^\dagger}{\trace\left(b^\dagger b \rho\right)} - \rho \right) \dd N_t^{\gamma_-} \, , \nonumber \\
	\mathcal{J}_{\gamma_+} (\rho) = & \gamma \bar{m} \Tr{b b^\dagger \rho} \rho \, \dd t + \left( \frac{b^\dagger \rho b}{\trace\left(b b^\dagger \rho\right)} - \rho \right) \dd N_t^{\gamma_+} \, ,
\end{align}
with weights
\begin{align} \label{Eq:Opto_Full_weights}
	\left< \dd N_t^\kappal \right> & = \kappa\, \Tr{a^\dagger a \rho} \dd t \, , \nonumber \\
	\left< \dd N_t^{\gamma_-} \right> & = \gamma \left(\bar{m} + 1 \right) \Tr{b^\dagger b \rho} \dd t \, , \nonumber \\
	\left< \dd N_t^{\gamma_+} \right> & = \gamma \bar{m}\, \Tr{b b^\dagger \rho} \dd t \, .
\end{align}
We emphasise again that the stochastic trajectory based only on physical observations must be computed using Eq.~(\ref{Eq:Opto_stochastic_observed}), while Eq.~(\ref{Eq:Opto_stochastic_full}) provides the full unravelling involving unobserved jumps that have to be averaged-over to restore the real dynamics.  Nevertheless, Eq.~(\ref{Eq:Opto_stochastic_full}) gives a pure evolution that is computationally simple to solve, so that it can be effectively used to generate click-patterns of detected photons, i.e.~the times at which $\dd N_t^\kappad=1$, by simply ignoring the records of other inaccessible jump-processes:~$\dd N_t^\kappal$, $\dd N_t^{\gamma_-}$, $\dd N_t^{\gamma_+}$.

\subsubsection{Linear regime} \label{Sec:Lin_opto}
If the coupling is weak compared to the driving strength of the system, a linearisation of the dynamics can be performed without losing much precision of the dynamics.  In particular, the optical and mechanical modes can be transformed into
\begin{subequations}
\begin{align}
	a \rightarrow & \, \alpha + \tilde{a} \\
	b \rightarrow & \, \beta + \tilde{b} \, ,
\end{align}
\end{subequations}
where $\alpha$ and $\beta$ are complex numbers and $\tilde{a}$ and $\tilde{b}$ are annihilation operators acting on a quantum fluctuation.  Applying the standard input-output formalism~\cite{Aspelmeyer2014,Liu2013}, one may derive the Langevin equations for both the classical:
\begin{subequations}
\label{Eq:Langevin_cl}
\begin{align}
	\dot{\alpha} = & \left(\mathrm{i} \Delta' - \frac{\kappa}{2} \right) \alpha - \frac{\mathrm{i} \Omega}{2} \\
	\dot{\beta} = & \left(-\mathrm{i} \omega_\mathrm{M} - \frac{\gamma}{2}\right) \beta - \mathrm{i} g | \alpha |^2,
\end{align}
\end{subequations}
and quantum:
\begin{subequations}
\label{Eq:Langevin_q}
\begin{align}
	\dot{\tilde{a}} = & \left(\mathrm{i} \Delta' - \frac{\kappa}{2} \right) \tilde{a} - \mathrm{i} g \alpha \left(\tilde{b} + \tilde{b}^\dagger \right) - \mathrm{i} g \tilde{a} \left(\tilde{b} + \tilde{b}^\dagger \right) \label{eq:opt_part}\\
	\dot{\tilde{b}} = &  \left(-\mathrm{i} \omega_\mathrm{M} - \frac{\gamma}{2}\right) \tilde{b} - \mathrm{i} g \left(\alpha^* \tilde{a} + \alpha \tilde{a}^\dagger \right) - \mathrm{i} g \tilde{a}^\dagger \tilde{a} - \sqrt{\gamma} \tilde{b}_\mathrm{in}, \label{eq:mech_part}
\end{align}
\end{subequations}
contributions to the modes with the effective detuning $\Delta' \coloneqq \Delta - g \left(\beta + \beta^* \right)$. Consistently with \eqnref{eq:master_optomech} no noise-terms appear in the optical part of quantum fluctuations \eref{eq:opt_part}, while thermal fluctuations are accounted for due to $\tilde{b}_\mathrm{in}$ present in the mechanics part \eref{eq:mech_part}.

For common parameter choices, i.e.~strong driving and weak coupling, the `classical part' is vastly greater than the `quantum part', i.e.~$|\alpha| \gg \langle \tilde{a} \rangle$ and $|\beta| \gg \langle \tilde{b} \rangle$.  In such a case, the nonlinear terms can be further dropped in the quantum part of Langevin equations \eref{Eq:Langevin_q}~\cite{Liu2013,Aspelmeyer2014}, what effectively corresponds to considering a linearised version of the Hamiltonian \eref{Eq:H_OM} that reads~\cite{Liu2013}:
\begin{align}
	H^\mathrm{lin} = & - \hbar \Delta' \tilde{a}^\dagger \tilde{a} + \hbar \omega_\mathrm{M} \tilde{b}^\dagger \tilde{b} + \left(G \tilde{a}^\dagger + G^* \tilde{a}\right)\left(\tilde{b} + \tilde{b}^\dagger \right) \, ,
	\label{eq:H_lin}
\end{align}
where $G \coloneqq \alpha g$.  

This Hamiltonian \eref{eq:H_lin} can be used then to describe the quantum part of the evolution, whereas the classical contributions to optical and mechanical modes, $\alpha$ and $\beta$, should be computed independently by solving equations \eref{Eq:Langevin_cl}. Typically, however, these attain very fast their \emph{steady-state} values, $\alpha_\mathrm{ss}$ and $\beta_\mathrm{ss}$, which can be always determined numerically. Hence, to facilitate the analysis when considering the optomechanical system in the linear regime in what follows, we shall always initialise both the optical and mechanical modes in their classical steady states. As a result, as the linear approximation requires the classical contribution to the modes to be much larger than quantum fluctuations, we expect both optical and mechanical modes to remain approximately in coherent states, $\ket{\alpha_\mathrm{ss}}$ and $\ket{\beta_\mathrm{ss}}$, at all times.

\begin{figure}[t]
	\centering
	\includegraphics[width=0.98\linewidth]{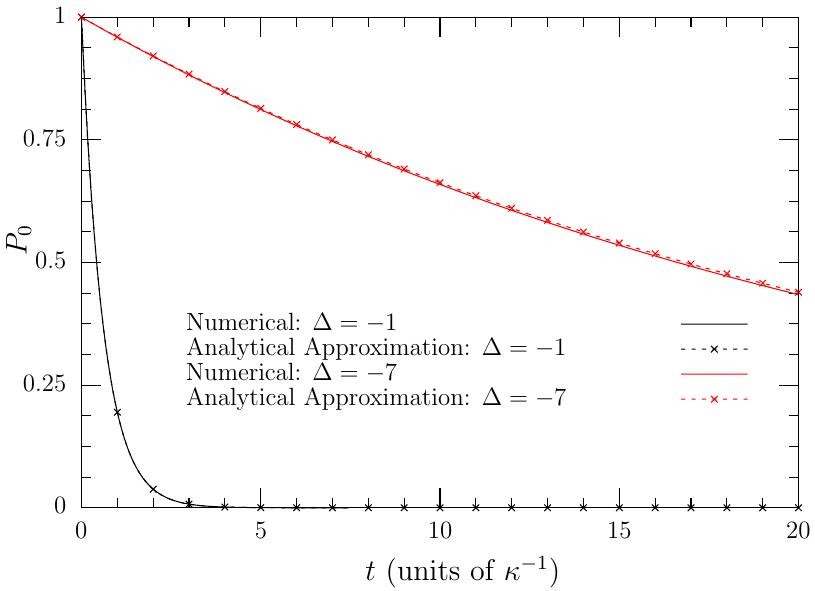}
	\caption{\textbf{Optomechanical system in the linear regime}. The probability of observing no photons, $P_0$, as a function of time (units of $\kappa^{-1}$) for two different values of detuning $\Delta$ (units of $\kappa$) computed using the full numeric solution (solid lines) and the analytic linearised approximation discarding the quantum fluctuations (dashed lines). In each case, the system is initialised in the stationary state.  The agreement between the two solutions proves the system to be approximately in a coherent state at all times and, as such, a Poissonian distribution for the number of photon-clicks over time is expected. In the above, other system parameters are set to $9\kappal = \kappad$ and $\bar{m} = 1$, $10g = \omega_\mathrm{M}/6 = \kappa$, while $\Omega = 2 \omega_\mathrm{M}$ and $\Gamma = \omega_\mathrm{M}/1200$.}
	\label{Fig:Linear_P0_test}
\end{figure}

We verify this intuition for the optical degree of freedom in Fig.~\ref{Fig:Linear_P0_test} by considering the probability of detecting no photons over time, $P_0(t)$.  On one hand, we solve numerically the exact dynamics \eref{eq:master_optomech} for the true Hamiltonian, $H^\mrm{OM}$ in \eqnref{Eq:H_OM}, in which case the no-photon probability is given by $P_0 (t) = \Tr{\rho_{0,\mathrm{acc}}(t)}$.  To perform the numerics, we take a truncated Fock space for both the optical and mechanical Hilbert spaces~\cite{Clark2022}. The truncation must not be too severe such that important dynamics are lost, but must result in a density matrix that is not too large to be computationally solvable.  On the other, we consider the approximate solution, i.e.~we neglect the quantum part of the dynamics \eref{Eq:Langevin_q} and assume only the classical contribution \eref{Eq:Langevin_cl}, in which case $P_0 (t) = \exp\left(- \kappad |\alpha(t)|^2 t\right)$. Moreover, as we take the system to initially be in the stationary state, \figref{Fig:Linear_P0_test} depicts $P_0(t) = \exp\left(-\kappad |\alpha_\mathrm{ss}|^2 t \right)$ that, indeed, up to negligible precision coincides with the exact numerical solution.

\subsubsection{Non-linear regime} \label{Sec:Non-lin_opto}
In a non-linear optomechanical system, we have stronger coupling, meaning the linearisation procedure highlighted previously is no longer valid.  As such, there are no analytic results available and numerical modelling is required, by again imposing a truncated Fock space and using the same methods ~\cite{Clark2022} as done previously to obtain the full numerical solution in the linear regime.  However, there is one key feature of such systems that we highlight and take advantage of in the following. In particular, we look at the energy spectrum of a non-linear optomechanical system.  This can be written as~\cite{Kronwald2013,Clark2022}
\begin{align} 
	\label{Eq:energy_levels}
	E(n_\trm{cav},n_\trm{mech}) = & - \hbar \Delta \,n_\trm{cav} + \hbar \omega_\mathrm{M} \,n_\trm{mech} \nonumber \\
	& - \hbar \frac{g^2}{\omega_\mathrm{M}} n_\trm{cav}^2 \, ,
\end{align}
where $n_\mathrm{cav}$ and $n_\mathrm{mech}$ are the occupation numbers of the cavity and mechanics.  Let us now consider what happens to the overall energy of the system whenever a photon is absorbed/emitted from the cavity.  In particular, we define special values of the detuning as:
\begin{align} \label{Eq:Detuning_n}
	\Delta_n \coloneqq - n g^2 / \omega_\mathrm{M} \, .
\end{align}
Then, with $n\in\mathbb{Z}^+$, the first and third terms in \eqnref{Eq:energy_levels} cancel for the cavity occupation number $n_\mathrm{cav}=n$. This means that transitioning from $n$ photons in the cavity to $0$ results in no energy change. As such, this transition is resonant and energetically favourable for the dynamics.

In this paper, we shall focus our work around two particular values of detuning for the non-linear system.  We take values for $n=1$ and $n=2$ in Eq.~(\ref{Eq:Detuning_n}).  In doing so, we create the resonant transition of the ground state with the first and second excited state respectively.  This creates interesting photon statistics, where for the $n=1$ case we see single photon emissions regularly, whereas for the $n=2$ case we see highly bunched light~\cite{Kronwald2013,Clark2022}.  These photon statistics clearly have very different properties, so thus should be distinguishable when trying to infer the value of the detuning.  We analyse the photon statistics further in the next subsection.

\subsection{Statistics of emitted photons}
\label{sec:click_stats}
\begin{figure}[t]
	\centering
	\includegraphics[width=0.98\linewidth]{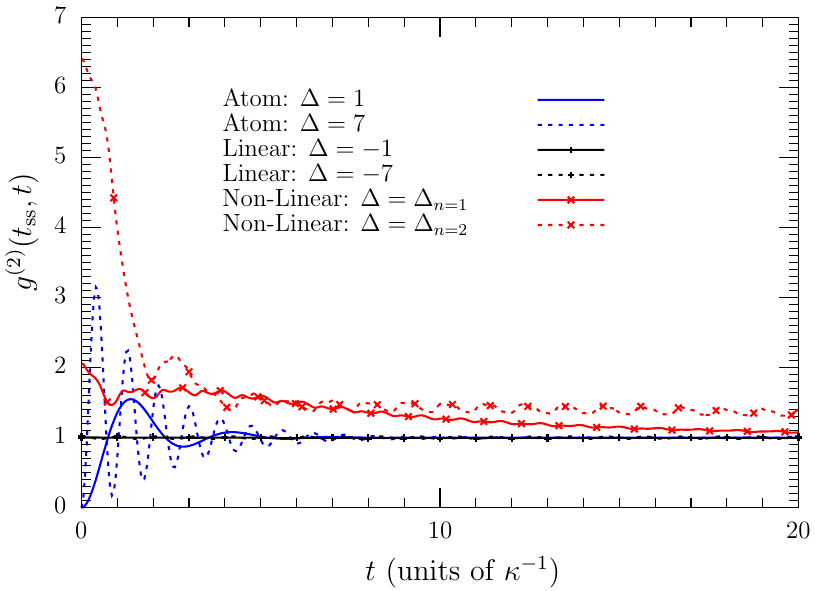}
	\caption{{\bf Stationary state $g^{(2)}$-function for the systems considered}.  For the two-level atom, oscillations are observed with their strength proportional to the detuning---the atom resets itself to the ground state and subsequently takes longer to reset for larger detunings. Nevertheless, the behaviour is not so complex.  For the linear optomechanical system, the behaviour is even more simple, as expected from the analysis in Sec.~\ref{Sec:Lin_opto}.  The $g^{(2)}$-function stays almost exactly at 1 for all time.  Finally, for the non-linear optomechanical system, we see the complex behaviour predicted from the bunching effects described in Sec.~\ref{Sec:Non-lin_opto} and in Refs.~\cite{Kronwald2013,Clark2022}.  The correlations may last for long times in such systems too, thus showing how complex the statistics in such a system may be, not being effectively Poissonian within the considered time interval.  For the two-level atom, we take the parameters as in Fig.~\ref{Fig:Waiting_atom_analytic} and for the linear optomechanical system we take the parameters as in Fig.~\ref{Fig:Linear_P0_test}.  The non-linear optomechanical system has parameters $9\kappal = \kappad$ and $\bar{m} = 1$, as for the linear system, but now $g/4 = \omega_\mathrm{M}/4\sqrt{2} = \kappa$ and $\Omega = 0.3 /\omega_\mathrm{M}$, $\Gamma = 10^{-3} \omega_\mathrm{M}$ for the remaining parameters, as in Refs.~\cite{Kronwald2013,Clark2022}.}
	\label{Fig:g2}
\end{figure}

We now analyse the behaviour of the photon statistics for both the two-level atom and the optomechanical system introduced above.  A natural way to analyse this is through the \emph{second-order correlation function}:
\begin{align}
	\label{Eq:g2}
	g^{(2)} (t_1 , t_2) := & \frac{p (t_1 , t_2)}{p(t_1) p(t_2)} = \frac{p(t_2 | t_1)}{p(t_2)} \, ,
\end{align}
where $t_2 \geq t_1$ and the probabilities are those of detecting a photon at the arguments. Typically, we consider $t_1$ to be a time in the stationary state, meaning $t_2$ can be merely thought of as a delay time $\tau_\dd$ after a first emission.  In such a case, the correlation function can be written as $g^{(2)} (\tau_\dd)$. As throughout this work, the unknown parameter to be inferred is the detuning of the driving, $\Delta$, we are interested in how the photon statistics vary as a function of $\Delta$. Hence, we depict in Fig.~\ref{Fig:g2} and discuss in more details below, the behaviours of the $g^{(2)}$-function for both systems and the relevant detuning values.

As well as the correlation functions, we shall also consider the average time to emit a fixed number of photons for each system.  This is a quantity that will clearly vary with the detuning and therefore would serve as a suitable summary statistic for later performing ABC.  The total time to emit $N$ photons can be expressed simply as the sum of the waiting time for $N$ photons, i.e.
\begin{align} \label{Eq:tN}
	t_N := & \sum\limits_{i=1}^{N} \Delta t_i \, ,
\end{align}
where $\Delta t_i$ is the time in between subsequent, $(i-1)$th and $i$th, photon emissions.

Considering any system that resets itself into a fixed state after an emission event, i.e.~a \emph{renewal process}, the expectation value for $t_N$ and its variance are  determined by the waiting-time distribution $w(\tau)$~\cite{Grimmett2001}. In particular, see App.~\ref{Sec:App_waiting}, it is straightforward to show that $\langle t_N \rangle = N \langle \Delta t \rangle$ and $\mathrm{Var} \, [t_N] = N \, \mathrm{Var} \, [\Delta t]$, where
\begin{subequations}
\label{Eq:Expectations_analytic}
\begin{align}
	\langle \Delta t \rangle = & \int\limits_0^\infty \dd \tau \, \tau\, w(\tau), \label{eq:av_w-time}\\
	\mathrm{Var} [\Delta t] = & \int\limits_0^\infty \dd \tau \, (\tau - \langle \Delta t \rangle)^2 \,w(\tau) \, \label{eq:var_w-time},
\end{align}
\end{subequations}
are the average waiting time for a single click and its variance, respectively.

It is convenient then to inspect the \emph{signal-to-noise ratio} (SNR) $\gamma$ that is defined via both \eref{eq:av_w-time}~and~\eref{eq:var_w-time} as~\cite{everitt2010cambridge}:
\begin{align}
	\gamma := & \frac{\langle \Delta t \rangle^2}{\mathrm{Var}[\Delta t]} \quad\underset{\trm{r.p.}}{=}\quad \frac{\langle t_N \rangle^2}{N\,\mathrm{Var}[t_N]},
	\label{eq:SNR}
\end{align}
which for a renewal process (r.p.) should also be recovered by considering a series of $N$ clicks and evaluating the expression above. Note that $\gamma$ is unity for any renewal process such that the variance of the average waiting time is equal to the square of its mean.  In particular, this is the case for a Poissonian system whose waiting-time distribution is exponential---a manifestation of the Markov property for the underlying stochastic process~\cite{Grimmett2001}.

\begin{figure*}[t]
	\centering
	\includegraphics[width=0.98\linewidth]{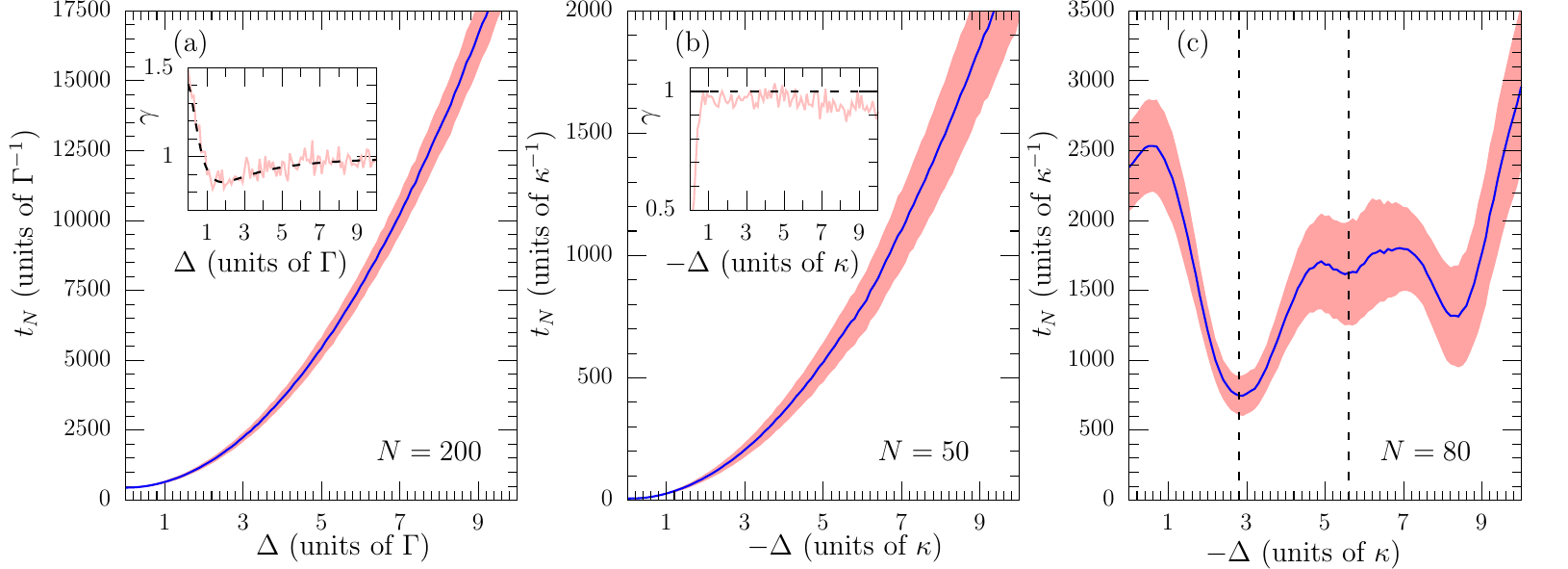}
	\caption{\textbf{Total time $t_N$ to emit a fixed number $N$ of photons} for each system considered, averaged over 2000 numerically generated trajectories for different values of detuning.  A monotonic behaviour for the two-level atom (a) and the linear optomechanical system (b) is observed, in contrast to a highly multivariate function for the non-linear system in (c), where the key detunings, $\Delta_{n=1}$ (left) and $\Delta_{n=2}$ (right), are highlighted with dashed vertical lines. The standard deviation of the samples trajectories is marked by the shaded area. The insets of (a) and (b) present the expected analytic behaviour (dark, dashed line) for the \emph{signal-to-noise ratio} (SNR), $\gamma$ defined in \eqnref{eq:SNR}, as a function of $\Delta$, as compared with the numerical prediction (r.h.s.~of \eqnref{eq:SNR}) obtained by averaging over the simulated $N$-click trajectories (feint, solid line). These indicate strong agreement between the numerics and theory, suggesting good reliability of the prepared library of trajectories.
	}
	\label{Fig:Avg_tN}
\end{figure*}

When evaluating the $t_N$-statistic \eref{Eq:tN} (and indeed simulating all trajectories of the systems), we initialise the system in a simple initial state, rather than the stationary state.  Nevertheless, due to the Ergodic principle, these two statistics are relevant, as all trajectories we consider will be averaged over long enough time, and we shall consider a large enough ensemble, so that all statistics are consistent.  This can also be thought of as allowing $N$ to be sufficiently large, such that almost the entirety of the clicks are independent of any initial state.  We show the expected value for the total time to emit $N$ photons in Fig.~\ref{Fig:Avg_tN}, based on averaging over the library of trajectories and also the analytic solutions, where available.  We now discuss each physical system individually.

\subsubsection{Two-level atom} \label{sec:2level_stats}
The two-level atom is the simplest of the systems that we consider.  The waiting-time distribution fully characterises when the next emission will occur because of the common resetting to the ground state.  We analyse this by averaging over a library of trajectories for two sample detunings in Figs.~\ref{Fig:Waiting_atom_analytic}(b-c).  In particular, we find that the trajectories reproduce the weighting time on average very well.  However, the variance increases with increasing value of detuning.  This is due to the flattening of the waiting-time distribution, which itself is caused by the reduction in oscillation between the ground and excited state at higher detunings.

In the case of a two-level atom, whenever a photon is emitted, the atom is reset into its ground state.  Such a process must yield correlations in the photon statistics, which are affected by the value of the detuning.  This is seen in the oscillations of the $g^{(2)}$-function in Fig.~\ref{Fig:g2}, which are stronger for the higher value of detuning.  Nevertheless, as the state resetting is common throughout the whole evolution, the waiting time and thus the intensity of emitted light is mostly characteristic of the data.

Moreover, using Eqs.~(\ref{eq:wait_time_two-level}) and (\ref{Eq:Expectations_analytic}), we calculate analytically the expected waiting time for the emission of $N$ photons and its variance, which respectively read \cite{Rinaldi2024}:
\begin{subequations}
\label{eq:w-time_stats}
\begin{align}
	\langle \Delta t \rangle_\mathrm{atom} = & \frac{\Gamma^2 +4 \Delta^2 + 2 \Omega^2}{\Gamma \Omega^2} \label{eq:av_w-time_atom}\, ,\\
	\mathrm{Var} [\Delta t]_\mathrm{atom} = &  \frac{\left(\Gamma^2 + 4 \Delta^2\right)^2 - 2\left(\Gamma^2 - 12 \Delta^2\right) \Omega^2 + 4 \Omega^2}{\Gamma^2 \Omega^4} \, . 
	\label{eq:var_w-time_atom}
\end{align}
\end{subequations}
The expression for the waiting time \eref{eq:av_w-time_atom} is compared with the one computed numerically using the library of trajectories in Fig.~\ref{Fig:Avg_tN}(a). We see good agreement between the two methodologies and, as one would expect, see a quadratic
growth as the detuning increases. Importantly, we therefore conclude that this quantity shall serve well as a summary statistic for ABC in the later sections, despite it ignoring the effect of correlations.

Similarly, using also the expression \eref{eq:var_w-time_atom}, we present the SNR \eref{eq:SNR} within the inset of Fig.~\ref{Fig:Avg_tN}(a). We see that $\gamma$ varies as a function of $\Delta$, which is to be expected for a renewal process with a waiting-time distribution that is \emph{not} exponential~\cite{Grimmett2001}. As such, the performance of ABC according to this statistic will not be the same for different values of the detuning.

\subsubsection{Linear optomechanical system} \label{sec:lin_optomech_stats}
The linear optomechanical system is more complex, but, as discussed in Sec.~\ref{Sec:Lin_opto}, the dynamics are to a good approximation one of a perturbed coherent state.  As such, upon reaching a stationary state, the system effectively no longer evolves, even under photon emission.  This is shown both by the agreement of our approximation with the numerical solution in Fig.~\ref{Fig:Linear_P0_test}, but also by the $g^{(2)}$-function in Fig.~\ref{Fig:g2}, where we see that the system remains approximately classical at all times (i.e.~$\forall \, \tau\ge0:\;g^{(2)}(\tau) \approx 1$), for both selected values of detuning.  As such, recalling \figref{Fig:Linear_P0_test}, the exponent of the exponential waiting-time distribution and, hence, the (inverse of) the average waiting time \eref{eq:av_w-time} contains all the information about the system.

We also see a monotonic increase in the total emission time $t_N$ in Fig.~\ref{Fig:Avg_tN}(b), thus acknowledging again the expected suitability of this statistic for ABC.  As the system approximately resets into a fixed state upon emission (as the dynamics are close to coherent), the $t_N$ and its variance should be almost exactly $N$ times the expressions \eref{Eq:Expectations_analytic}. Still, as we do not have the exact form of $w(t)$ here, we cannot provide an analytic solution. Unlike the two-level atom, however, we have an (approximately) coherent-light emitter with the waiting time of photon-click events following (approximately) an exponential distribution. This implies that the SNR \eref{eq:SNR} should be close to unity irrespectively of the value of detuning, as verified within the inset of Fig.~\ref{Fig:Avg_tN}(b).

\subsubsection{Non-linear optomechanical system} \label{sec:nonlin_optomech_stats}
The non-linear system is far more complex than the two previous examples. Not only is there no analytic solution, but also the system cannot even be approximately assumed to reset itself to a common state upon emission. Moreover, the effect of a jump is highly non-trivial, resulting in correlated photon statistics. The reason for this can be seen by analysing Eq.~(\ref{Eq:energy_levels}) more carefully.

When the dynamics are resonant, i.e.~in the $n=1,2$ regimes described previously, the system favours transitioning to and from the ground state and the $n^\mathrm{th}$ level in the cavity.  For the $n=1$ case, this means that the optical part of the dynamics will favour emitting single photons, effectively acting as a two-level atom.  For the $n=2$ case, the system will typically emit photons in pairs.  This is due to the cavity rapidly decaying from the $2$-photon state to the ground state.  This results in a noticeably high $g^{(2)}$-function, seen in Fig.~\ref{Fig:g2}, due to this strong bunching effect.  These correlations are significant in the photon statistics and should therefore be taken into account to fully characterise the emission profile.

This is further shown in the total emission time in Fig.~\ref{Fig:Avg_tN}(c), where we see a non-monotonic function in detuning.  We therefore expect this will limit the performance of ABC utilising this statistic.  Nevertheless, there is a well-isolated minimum around $\Delta_{n=1}$ that, when approached from above or below, should still provide sufficient information about the detuning $\Delta$ by inspecting the total emission time $t_N$. However, note that the observed variance $\mathrm{Var}[t_N]$ is non-zero at $\Delta=\Delta_{n=1}$, so we expect the $t_N$-statistic to be not as informative as in the case of two-level atom or linear optomechanics around $\Delta=0$.

\section{Parameter inference and the application of the ABC algorithm} 
\label{Sec:ABC}

In this section, we now apply the ABC algorithm to the exemplary systems analysed in the previous section.  For each system, we first identify the summary statistics used, before then verifying the performance of the algorithm against the true posterior obtained by analytical or numerical methods.  As stated previously, we take the detuning $\Delta$ to be the unknown parameter throughout, over which we are to obtain the posterior. For all systems we perform ABC with summary statistics wished to be sufficient. In the case of the non-linear optomechanics, however, we first choose total time as a summary statistic to verify its limitations, before then making ABC effective by choosing then the histogram of waiting times as a summary statistic, which is importantly capable of certifying photon-bunching (quantum) effects.

\subsection{Benchmarking the performance of ABC}
\subsubsection{Prior choice}
In what follows, we take a flat prior distribution across the parameter space.  In doing so, we ensure that any results we obtain are not a consequence of making preference in the choice of the true parameter.  Moreover, as the ABC algorithm is affected by the prior as the proposal distribution, such choice ensures that we are not just reproducing the posterior by only selecting samples from the correct region.

\subsubsection{Quantifying the error of ABC}
To quantify how well the ABC algorithm performs, we compute the \emph{fidelity} of the posteriors it produces by evaluating the relevant Bhattacharyya coefficient: 
\begin{align} \label{Eq:Fidelity}
	F[p_\mathrm{post},p_\mathrm{ABC}|D_t] \coloneqq & \int \dd \Delta \sqrt{p_\mathrm{post}(\Delta | D_t)} \sqrt{p_\mathrm{ABC}(\Delta | D_t)} \,,
\end{align}
where $p_\mathrm{post}(\Delta | D_t)$ denotes the true posterior, obtained through analytical means for the two-level atom or numerically for the optomechanical system, and $p_\mathrm{ABC}(\Delta | D_t)$ is the posterior obtained through ABC. We show the results of this calculation in the insets of main plots (depicting posteriors) as a function of the \emph{number of samples used} $\nu$ (recall the ABC-algorithm~\ref{alg:ABC}), \emph{not} the number of samples accepted.  If the summary statistics are sufficient, then in the limit of an infinitely tight bound with infinitely many samples, the fidelity averaged over many trajectories should converge to 1.  This is due to the definition of sufficiency in Eq.~(\ref{Eq:Sufficient}), meaning in such a case we are effectively sampling directly from the posterior \cite{Didelot2011}.  Of course with realistic numerical capabilities, it is not possible to do this, but for reasonably tight bounds and our finite library, we expect to come close to unity, if the summary statistics are chosen well.

\begin{figure*}[t]
	\centering
	\includegraphics[width=0.90\linewidth]{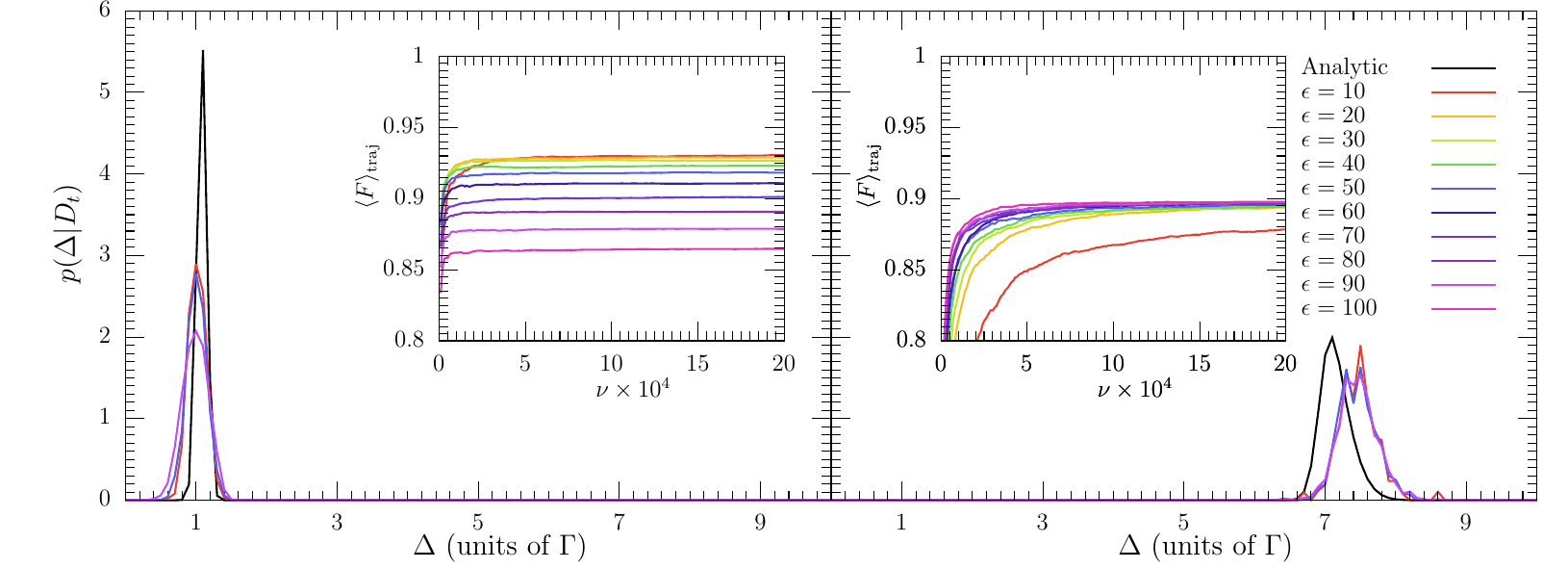}
	\caption{\textbf{ABC performance for the two-level atom with total time} to emit $N=200$ photons used as the summary statistic, for two different values of detuning:~$\Delta = 1 \Gamma$ (left) and $\Delta = 7 \Gamma$ (right).  The main plots depict the ABC-based posteriors (in colour) for different values of the acceptance threshold $\epsilon$, which become closer and closer to the true posterior (black) as the threshold is decreased, so long as a sufficiently large number of samples is used.  All the posteriors shown are for one implementation of ABC algorithm on a specific photon-click trajectory $D_t$.  However, we test this further by averaging over 20 different trajectories (and 10 different implementations of ABC) for both values of $\Delta$, and show the average fidelity $\langle F\rangle_\textrm{traj}$ as a function of the number of samples used, $\nu$, within the insets. These demonstrate that the ABC algorithm quickly reaches its performance limit for only a relatively small sample numbers. System parameters are taken as in Fig.~\ref{Fig:Waiting_atom_analytic}.
	}
	\label{Fig:Post_atom}
\end{figure*}

For the two-level atom, as we are able to analyse trajectories analytically, we can produce analytic posteriors with ease. In such a case, to determine the average fidelity of ABC, we analyse its performance by averaging not only over multiple implementations of the ABC algorithm (considering also different thresholds $\epsilon$ in \eqnref{Eq:Accept_ABC}) for a given photon-click trajectory, but also over distinct trajectories being observed. We do this for two different true values of the detuning $\Delta$, to demonstrate the performance of the algorithm across the parameter range.  For the optomechanical system, however, obtaining the true posterior is highly expensive computationally.  Hence, instead of averaging over multiple trajectories, we take three exemplary trajectories for each of the two different true values of $\Delta$, and evaluate the ABC performance upon averaging only over multiple implementations of ABC (again for different values of the threshold $\epsilon$).

\subsubsection{Optimising the ABC algorithm}
The performance of the ABC algorithms, and hence the fidelity $F$ introduced in Eq.~(\ref{Eq:Fidelity}), strongly depends on the choice of threshold $\epsilon$ in \eqnref{Eq:Accept_ABC} given the number of samples $\nu$ considered. As more and more samples are used, the fidelity gradually plateaus and reaches a steady value.  One may naively expect that the tighter the threshold $\epsilon$, the better performance of the algorithm. However, a cost of this is an increased number of samples $\nu$ required to reach this plateau. We demonstrate this for all of our fidelity plots in what follows, and indeed observe this behaviour.  In some cases, we see that the tightest choice of $\epsilon$ is not optimal, as it results in undersampling even for the maximal possible $\nu$ considered.  In most cases, however, we observe that, given the large number of trajectories we possess in the library, the use of a big number of samples results actually in oversampling. For the two-level atom and linear optomechanical system, we evaluate the performance up to $\nu = 2\times 10^5$, whereas for the non-linear system we use up to $\nu = 4 \times 10^5$. In this regard, our results allow us to optimise the ABC algorithm in each case by identifying the number of samples $\nu$ required, given a particular value of threshold $\epsilon$.

\subsection{Two-level atom}
For the two-level atom, the posterior may be obtained analytically, so that the use of the ABC algorithm is in principle not advantageous. However, as we are able to obtain analytic posteriors cheaply, we can benchmark the performance of ABC by applying it to a large number of true posteriors straightforwardly.

\subsubsection{Summary statistic: total time}
Recalling \secref{sec:2level_stats}, a simple summary statistic should be enough to reproduce the posterior due to the emission profile constituting a renewal process. In particular, the \emph{total time} to emit a certain number of photons, $t_N$ in \eqnref{Eq:tN}, should be reasonably characteristic of any dataset. It is also a monotonic function of $\Delta$, recall \figref{Fig:Avg_tN}(a), and should thus be able to distinguish well its different values. For the two-level atom, we take the number of emitted photons to be $N=200$, so that our summary statistic is $S(D_t) = t_{N=200}(D_t)$.

\begin{figure*}[t]
	\centering
	\includegraphics[width=0.92\linewidth]{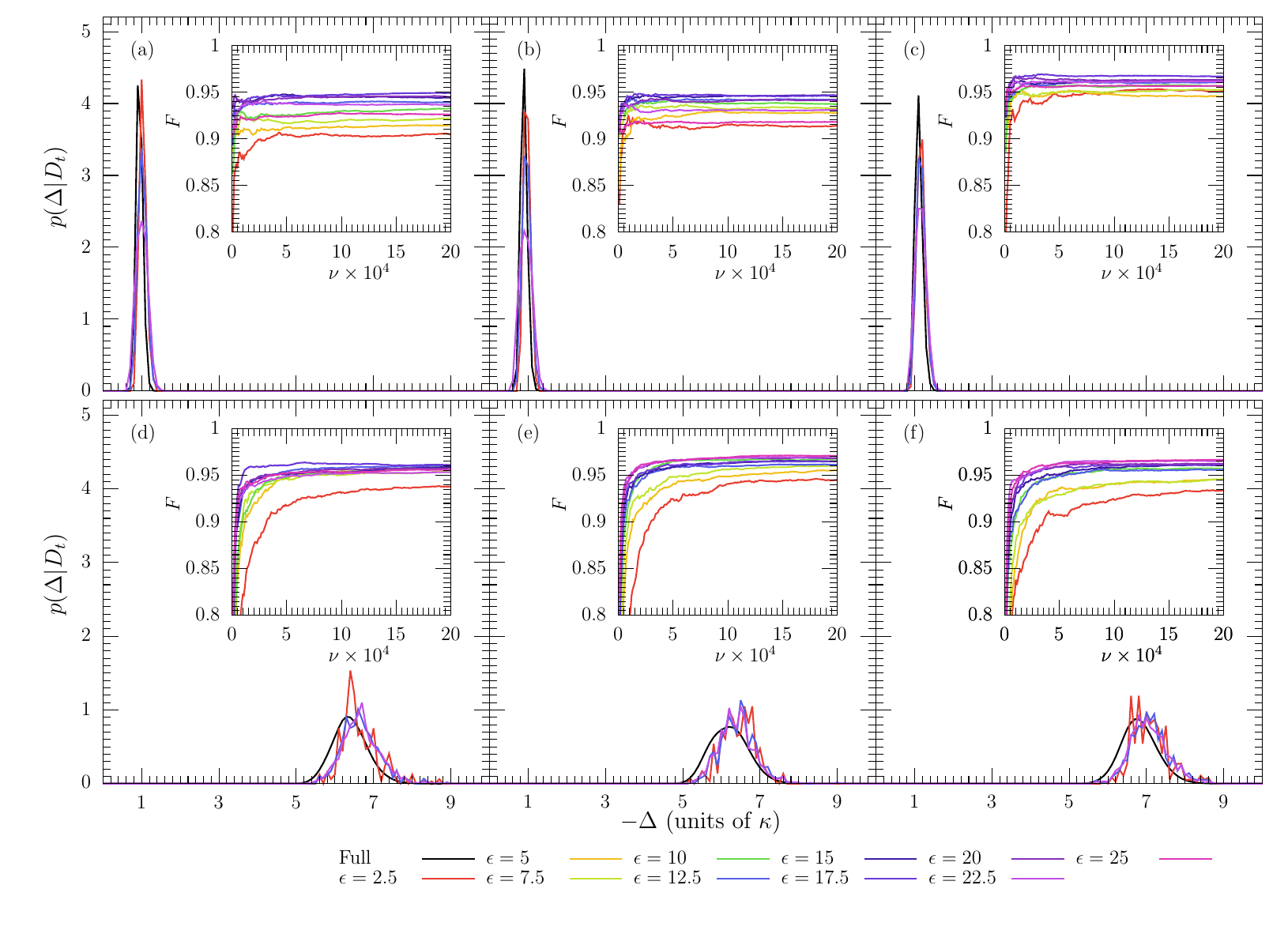}
	\caption{\textbf{ABC performance for the linear optomechanical system with total time}	to emit $N=50$ photons used as the summary statistic, for two different values of detuning:~$\Delta = -1 \kappa$ (a-c) and $\Delta = -7 \kappa$ (d-f).  In each plot the ABC-based posterior distributions (different colours for different thresholds $\epsilon$) are presented against the true posterior (black). As the latter is obtained by performing brute-force numerics, this precludes from presenting the performance ABC averaged over different photon-click trajectories. That is why, the insets present (in contrast to Fig.~\ref{Fig:Post_atom}) the fidelity \eref{Eq:Fidelity}, $F$, for each of the six different photon-click trajectories $D_t$ considered (still averaged over 10 implementations of ABC for each threshold). We find good agreement with the true posterior in all cases, demonstrating the validity of ABC.  Moreover, the performance is satisfactory independently of detuning value---being in line with our understanding of photon-click statistics:~the total time and its signal-to-noise ratio (SNR) presented in \figref{Fig:Avg_tN}(b).
	}
	\label{Fig:lin_samp}
\end{figure*}

\subsubsection*{Results}
Using this summary statistic, we apply the ABC algorithm and show the results in Fig.~\ref{Fig:Post_atom} for a range of different acceptance thresholds $\epsilon$.  We see for two different true values of the detuning, $\Delta = 1,7$ ($\times\Gamma$), the algorithm is capable of reproducing the true posterior to a good approximation.  For $\Delta = 1$, the algorithm saturates its potential very quickly and to a high fidelity, with looser thresholds doing so faster, but reaching worse overall fidelity.  For $\Delta = 7$, some errors start to appear, due to the flattening of the weighting-time distribution---recall \figref{Fig:Waiting_atom_analytic}(c)---which is then less sensitive to variations of $\Delta$. Although this indicates that more sophisticated statistics accounting for multi-time correlations are necessary to achieve better performance at $\Delta=7$, the ABC algorithm still works reasonably well.

In the insets of Fig.~\ref{Fig:Post_atom} we show the average fidelity (\ref{Eq:Fidelity}) between the posteriors generated through ABC and the respective true ones.  We average over 20 different trajectories (i.e.~true posteriors) and 10 unique implementations of ABC for each trajectory, to show the average performance of ABC.  We see clearly that the number of samples $\nu$ required to saturate the performance of ABC is greater for $\Delta = 7$ than $\Delta = 1$.  What we also see, within the range of $\nu$ considered, is that the performance is saturated for all but the tightest $\epsilon$-threshold value.  However, for $\Delta = 7$ there is hardly any improvement in the maximal average fidelity attained as $\epsilon$ is reduced. This is a consequence of trajectories possessing a much longer total time---recall \figref{Fig:Avg_tN}(a) where $\langle t_N\rangle\approx500(10000)$ for $\Delta=1(7)$. Hence, as it is the difference between total times that is used for acceptance, for thresholds to have the same significance these should be rescaled by a factor of $\approx20$. However, we must avoid any adjustments of the ABC algorithm depending on the $\Delta$-value, as these would require the estimated parameter to be known in advance. Hence, we are forced to select a value of $\epsilon$ that to the best extent is tight enough within the whole parameter range $\Delta\in[0,10]$ considered---for example, from \figref{Fig:Post_atom} we may infer $\epsilon = 20 \kappa^{-1}$ as an ideal candidate for all-round performance (given $\nu=20\times10^4$ samples).

Lastly, let us comment on the bias exhibited by the posteriors in \figref{Fig:Post_atom}, which naively seem to be shifted to the left(right) for $\Delta = 1(7)$.  This, however, is a standard artefact of Bayesian inference and inspecting performance based only on a single trajectory. Moreover, as our summary statistic based on the total time is not exactly sufficient, it may also introduce some systematic errors. Nevertheless, when multiple trajectories are analysed, e.g.~to compute the average fidelities in Fig.~\ref{Fig:Post_atom}, we observe that the bias is washed out on average.

\begin{figure*}[t]
	\centering
	\includegraphics[width=0.92\linewidth]{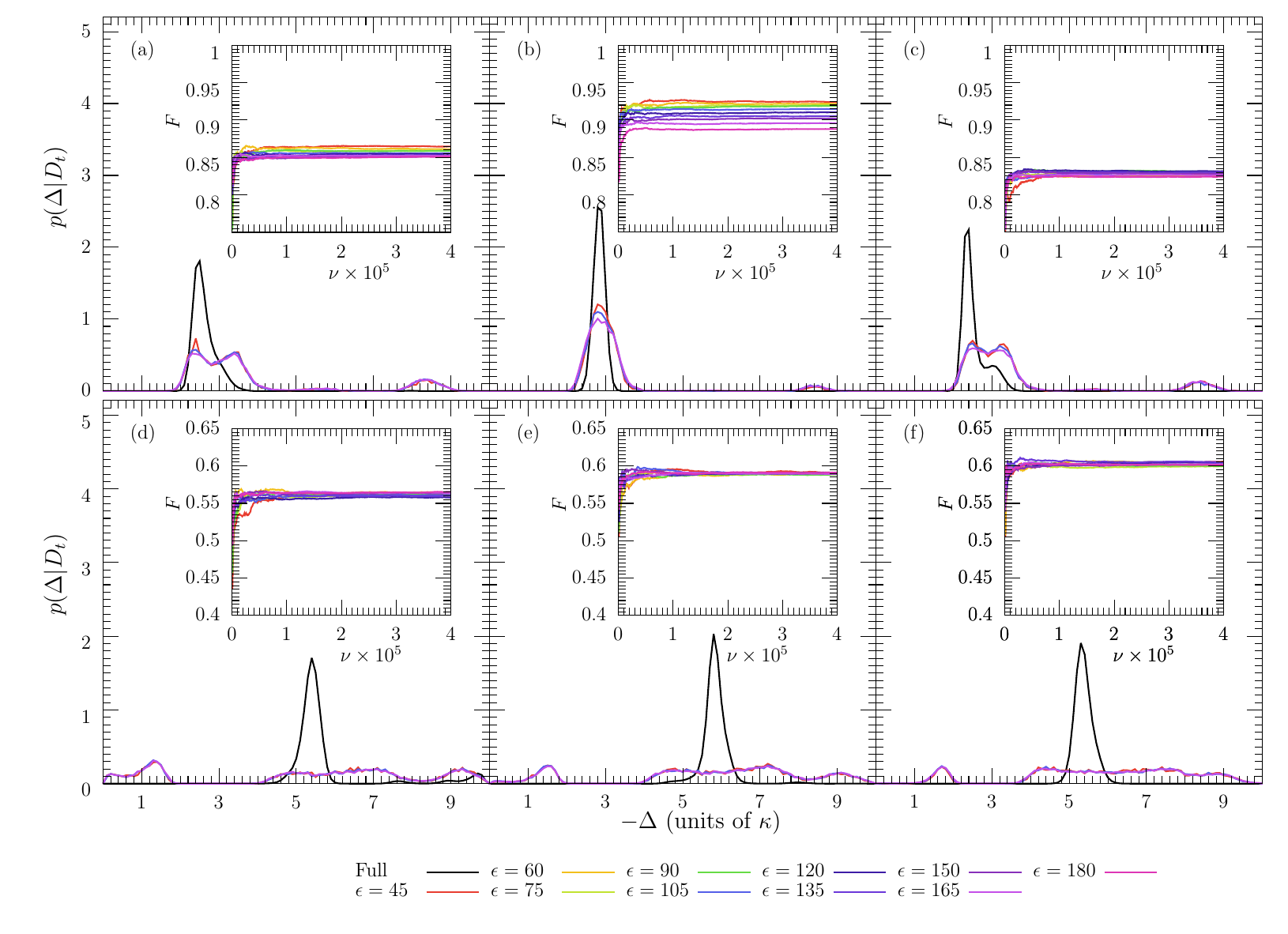}
	\caption{\textbf{ABC performance for the non-linear optomechanical system with total time} to emit $N=80$ photons used as the summary statistic, for two different values of detuning:~$\Delta = \Delta_{n=1} \approx -2.8 \kappa$ (a-c) and $\Delta = \Delta_{n=2} \approx -5.6 \kappa$ (d-f). In each subplot (a-f) a different photon-click trajectory is considered and, as before, ABC-based posterior distributions (different colours for different thresholds $\epsilon$) are presented against the true posterior (black). In (a)-(c) we see a reasonable performance of ABC, though clearly less accurate than in the case of linear system.  In (d)-(f), however, the algorithm fails to reproduce the posterior in all the cases, resulting in low fidelity (averaged over 10 ABC implementations) within each inset. As expected from \figref{Fig:Avg_tN}(c), the total time $t_N$ can now no longer be considered as a sufficient summary statistic, especially around $\Delta = \Delta_{n=2}$.
	}
	\label{Fig:non_lin_samp_1}
\end{figure*}

\subsection{Linear optomechanical system}
The linear optomechanical system requires numerical solutions to identify the true posterior distribution. Thus, the ABC algorithm provides a useful speed-up. Again, we discuss first the summary statistics employed before presenting the performance of the ABC algorithm.

\subsubsection{Summary statistic: total time}
In \secref{sec:lin_optomech_stats}, it was shown that the intensity of emitted light, i.e.~the exponent of the exponential waiting-time distribution, should be almost completely characteristic of the emission profile for a linear optomechanical system.  Hence, the \emph{total time} for a fixed number of photon emissions, $t_N$ in \eqnref{Eq:tN}, should again be sufficient as a summary statistic.
However, as the simulation of dynamics is more computationally exhaustive now, we restrict ourselves to 50 photons, i.e.~$S(D_t) = t_{N=50}(D_t)$.  This is (at least approximately) sufficient due to the simplicity of the exponential average waiting-time distribution---assured by the (almost) Poissonian emission profile discussed in \secref{sec:lin_optomech_stats}.

\subsubsection*{Results}
We show the results of implementing ABC with a variety of thresholds for the linear optomechanical system in Fig.~\ref{Fig:lin_samp}.  We see strong agreement between implementing ABC for the linear optomechanical system, even for larger detuning in this case.  This is in line with our expectations for this summary statistic and the linear optomechanical system, due to the simplistic nature of the photon emission patterns forming approximately an exponential distribution, whose exponent is monotonic in the detuning parameter---recall \figref{Fig:Avg_tN}(b).

We again quantify the performance of the algorithm through the fidelity \eref{Eq:Fidelity}. However, unlike the two-level atom, obtaining true posteriors requires numerical solutions to the dynamics, which are extremely time consuming.  We instead show the results of three indicative trajectories for $\Delta = -1 \kappa$ in Fig.~\ref{Fig:lin_samp}(a)-(c) and for $\Delta = -7 \kappa$ in Fig.~\ref{Fig:lin_samp}(d)-(f), while the insets now show just the fidelity averaged over 10 implementations of the ABC algorithm---being no longer averaged over different photon-click trajectories.  Nevertheless, we see reasonable agreement for all the trajectories considered, suggesting the performance is relatively consistent.

In particular, we also see consistent performance for the different values of detuning.  This is in agreement with expectations from the SNR $\gamma$ presented in the inset of \figref{Fig:Avg_tN}(b), which is mostly invariant to changes of the detuning.  Moreover, as the statistics form an exponential distribution, the emitted light is almost coherent with no inter-photon correlations, $\forall\tau:\;g^{(2)}(\tau)\approx1$ in \figref{Fig:g2}, meaning that the total time for a fixed number of emissions \eref{Eq:tN} is highly informative. We therefore see an improved performance compared to the two-level atom.
	
Contrastingly, we see that choosing the tightest threshold in the ABC algorithm, $\epsilon=2.5$ (red) in \figref{Fig:lin_samp}, does not yield the optimal performance.  This can be explained by the fact that we consider individual trajectories, rather than an average over many.  Thus, our trajectories could be unrepresentative, in that their summary statistics are not represented well enough by the samples in our library. Therefore, we need less tight thresholds to capture their behaviour, unless the number of samples $\nu$ can be significantly increased. Nevertheless, comparing the overall shapes of posteriors in \figref{Fig:lin_samp}, the fact of choosing too tight thresholds is relatively negligible.

\subsection{Non-linear optomechanical system}
Finally, we consider the non-linear optomechanical system.  As already emphasised, the dynamics is now complex, resulting in non-trivial photon statistics. Nevertheless, we first proceed with the same summary statistics of the total emission time, in order to explicitly demonstrate that, as expected, it will \emph{not} be satisfactory, especially for detunings yielding photon-bunching effects.

\subsubsection{Summary statistic: total time}
We consider again the time for a fixed number of photon emissions, $t_N$ in \eqnref{Eq:tN}, as the summary statistic, but increase the number of detection events to 80, meaning $S(D_t) = t_{N=80}(D_t)$.  Nonetheless, as $t_N$ is highly non-monotonic in detuning $\Delta$, recall Fig.~\ref{Fig:Avg_tN}(c), we expect such an approach to generally fail. However, as the global minimum in Fig.~\ref{Fig:Avg_tN} occurs at $\Delta=\Delta_{n=1}$, we also expect $\Delta$ to be still inferable around that point. Hence, in order to also confirm this behaviour, we still examine the performance of ABC with $t_N$ used as the only summary statistic for the non-linear optomechanical system.

\subsubsection*{Results}
We show the results compared to true posteriors obtained by numerical means in Fig.~\ref{Fig:non_lin_samp_1}, as for the linear optomechanical system.  For Fig.~\ref{Fig:non_lin_samp_1}(a)-(c) we set $\Delta = \Delta_{n=1}$, while for Fig.~\ref{Fig:non_lin_samp_1}(d)-(f) we set $\Delta = \Delta_{n=2}$.  As expected, in the $n=1$ regime, we see a good performance of ABC, capable of identifying the parameter range around $\Delta_{n=1}$ to a high degree of accuracy, while mistaking it slightly for $\Delta_{n=3}=-8.4\kappa$ at which the most similar total emission times can occur---recall \figref{Fig:Avg_tN}(c). For the $n=2$ regime, however, the algorithm does not reproduce the posterior, being capable only to \emph{eliminate} the ``inferable'' values of the detuning around $\Delta_{n=1}$ from the distribution.

For the $n=1$ regime, we see that the tighter thresholds achieve better fidelities than the looser ones, meaning, unlike for the linear system, the results are inline with expectations.  Moreover, they achieve this extremely quickly.  For the $n=2$ case, there is little difference in the performance of the thresholds, as there are never meaningful results, even when allowed to run for a large number of samples.  To certify this, we run the algorithm over twice as many samples ($\nu_\textrm{max} = 4 \times 10^5$), in order to certify that the fidelity has indeed saturated, see the insets of \figref{Fig:non_lin_samp_1}(d-f).

\begin{figure}[t]
	\centering
	\includegraphics[width=0.9\columnwidth]{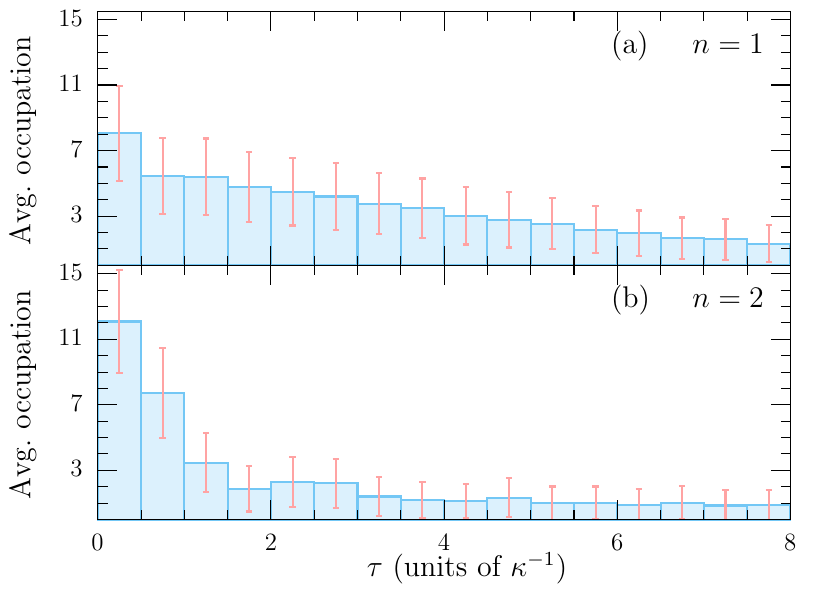}
	\caption{\textbf{Histogram of waiting times $\tau$ in between subsequent photon-clicks} for detuning $\Delta_{n=1}$ in (a) and $\Delta_{n=2}$ in (b). In each case, the data represents average occupancies for 2000 trajectories, each of $N=80$ photo-clicks. By focussing on the \emph{short-time regime} in the above ($\tau\le8\kappa^{-1}$), we see a clear difference between the distributions:~(a) -- an exponential decreasing modestly due to uncorrelated single-photon ($n=1$) emissions being dominant;~versus (b) -- a distribution dropping steeply at low $\tau$ due to double-photon ($n=2$) emissions being favoured. The error bars mark the standard deviation of the trajectory-samples used.}
\label{Fig:non_lin_avg}
\end{figure}

\begin{figure*}[t]
	\centering
	\includegraphics[width=0.92\linewidth]{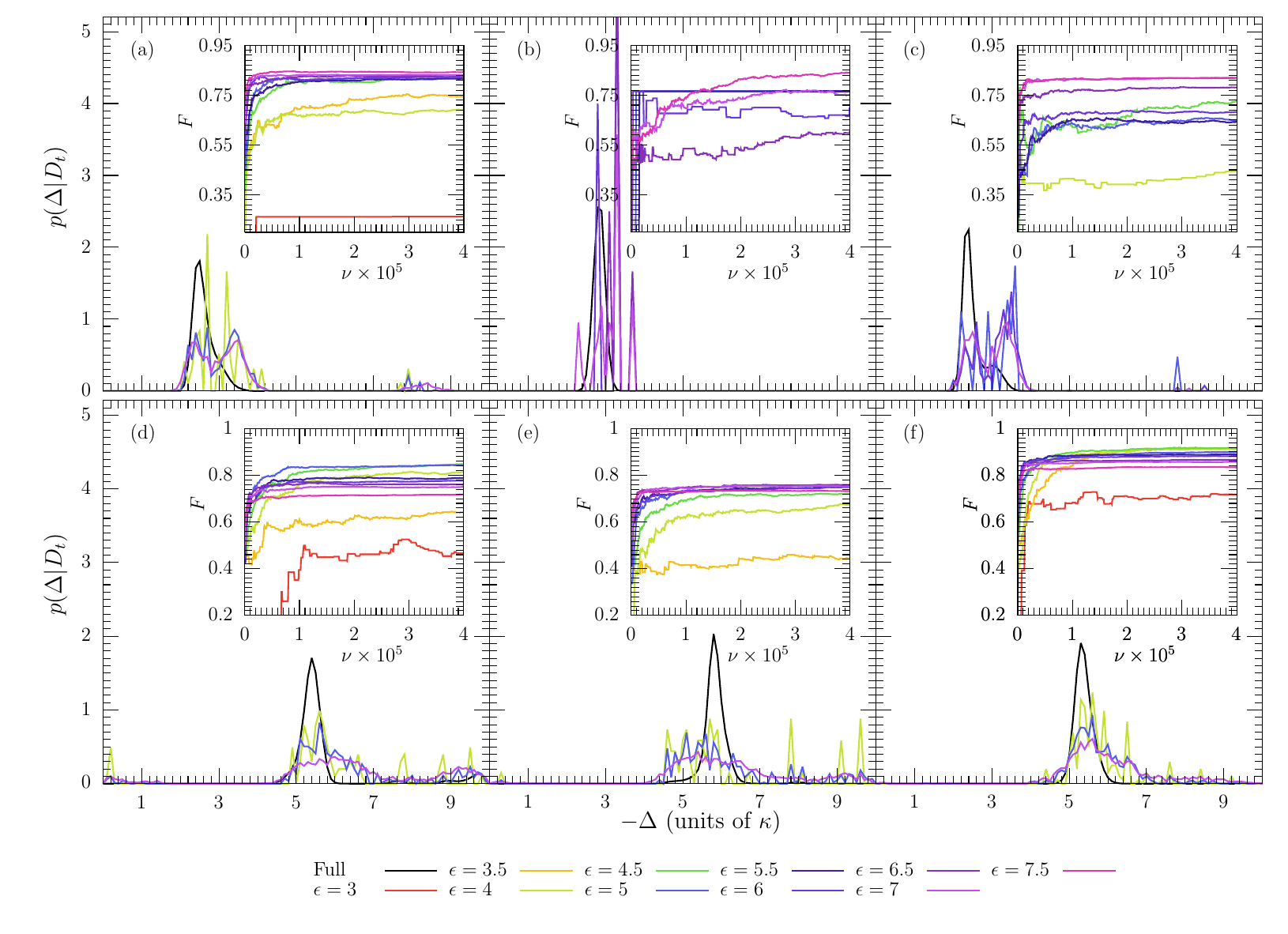}
	\caption{
	\textbf{ABC performance for the non-linear optomechanical system employing the (truncated) histogram of waiting times} as the summary statistic. In particular, the acceptance criterion of the ABC utilises the L2-norm \eref{eq:L2norm} between the histograms representing the waiting-time distribution in the range of short-time intervals, with each trajectory consisting of $N=80$ photon-clicks. As in \figref{Fig:non_lin_samp_1}, two different values of detuning are considered:~$\Delta = \Delta_{n=1} \approx -2.8 \kappa$ (a-c) and $\Delta = \Delta_{n=2} \approx -5.6 \kappa$ (d-f), with each subplot (a-f) being computed for a different photon-click trajectory. Importantly, in all the cases the ABC algorithms is now capable of reproducing the true posterior (black), while a less stringent $\epsilon$-threshold (different colours) must now be chosen due to much higher numbers of samples, $\nu$, being required for the method to converge.
	}
	\label{Fig:non-lin_samp_2_distance}
\end{figure*}

\subsubsection{Summary statistic: histogram of waiting times}
As confirmed, the total emission time does not constitute a sufficient summary statistic for non-linear optomechanics, as it is monotonic only within restricted ranges of the detuning parameter $\Delta$, see \figref{Fig:Avg_tN}(c), but it is also incapable of distinguishing distinct types of inter-photon correlations. However, before proposing other summary statistics, one could consider a ``full-data'' approach~\cite{Drovandi2022} with the acceptance rule \eref{Eq:Accept_ABC} being based directly on the comparison between the observed $D_t$ and sampled $D'_t$ photon-click trajectories. Unfortunately, \eqnref{Eq:Accept_ABC} would then need to resort to a temporal distance measure~\cite{Su2020}, $\delta(D_t,D'_t)$, applicable to point processes~\cite{everitt2010cambridge}, which are known, e.g.~the Wasserstein distance~\cite{Xiao2017}, to be computationally hard, not to mention calibrating then the $\epsilon$-threshold that works within the whole range of $\Delta$ considered.  Moreover, the library used within the ABC algorithm would then need to contain the actual trajectories $D'_t$, as storing summary statistics $S(D'_t)$ labelled by $\Delta$-parameter values would no longer be enough.

Because of these challenges, we propose to use the (truncated) \emph{histogram of waiting times} in between subsequent photon clicks as a summary statistic, which is known to capture the photon bunching and anti-bunching effects~\cite{Kronwald2013,Clark2022} and therefore should be sufficient to discriminate detuning values within the range covering values $\Delta_{n=1}$ and $\Delta_{n=2}$ marked within \figref{Fig:Avg_tN}(c). This is confirmed by Fig.~\ref{Fig:non_lin_avg}, in which we present the so-obtained waiting-time distributions built upon considering 2000 simulated trajectories, each one containing $N=80$ photon-clicks, for the two detuning values of interest:~$\Delta_{n=1}$ and $\Delta_{n=2}$.

Although a histogram of waiting times does not contain the full information about a given trajectory---in particular, the ordering of the clicks is lost, so that we expect it to be insufficient around $\Delta_{n=3}$ with photons being then emitted in triplets---we show that it suffices for our purposes. What is more, we ignore the non-stationarity of the underlying optomechanical dynamics, as it occurs at negligible timescales when compared to the total duration times of the trajectories considered. 

Importantly, when constructing the histogram we restrict ourselves to inter-click intervals of short duration,  i.e.~$\tau<8\kappa^{-1}$ in \figref{Fig:non_lin_avg}, as we expect the photon bunching effects to alter the---otherwise exponential---waiting-time distribution only at short timescales. Although we lose in this way the information about the total emission time that can be inferred only from the full histogram, we presume it to be (mostly) irrelevant and decide not to include (again) the total time as a secondary summary statistic. 

Finally, within the ABC algorithm \ref{alg:ABC} we employ the L2-norm within the acceptance rule \eref{Eq:Accept_ABC}, i.e.:
\begin{align}
	\delta(S(D_t),S(D'_t)) = & \sqrt{\sum\limits_i \left(x_i - x'_i \right)^2} \, ,
	\label{eq:L2norm}
\end{align}
where $S(D_t)$ is the (truncated) time-binned histogram of the inter-click waiting times as in \figref{Fig:non_lin_avg}, and the $x_i$'s are the respective sizes of histogram entries.

\subsubsection*{Results}
In Fig.~\ref{Fig:non-lin_samp_2_distance}, we present the resulting ABC-based posterior distributions, which should be compared directly against the previous ones of \figref{Fig:non_lin_samp_1} with the same photon-click trajectories being considered.  Crucially, the ABC algorithm is now capable of reproducing the true posterior in both the $n=1$ and $n=2$ regimes. For the $n=1$ regime, we see in Fig.~\ref{Fig:non-lin_samp_2_distance}(a-c) better performance of the ABC than in \figref{Fig:non_lin_samp_1}(a-c) as the values of $\Delta$ around $n=3$ are now excluded within the posterior.  However, the distribution itself is no longer that smooth. The latter fact is a consequence of the discretisation induced by the finite width of bins used to construct the histogram and a low number of clicks ($N=80$) composing single trajectories for which the L2-norm \eref{eq:L2norm} is evaluated. Still, despite similar discontinuities we observe for $n=2$ a significant improvement in \figref{Fig:non-lin_samp_2_distance}(a-f). As we tighten the $\epsilon$-thresholds in \figref{Fig:non-lin_samp_2_distance}, we observe that none of the samples is sometimes accepted---indicating that these would require a much higher number of samples than $\nu=4\times10^{5}$ here used. Nevertheless, for looser thresholds we manage to obtain good results even with the number of samples at hand.

We also emphasise that it should be possible to improve the ABC-method further with better calibration, not only of the acceptance threshold, but also by optimising the width of time-bins considered to construct the histograms in \figref{Fig:non_lin_avg}. In principle, such a calibration offers more flexibility than when comparing only the summary statistics. However, even with the minimal calibration performed here, we see already a good overall performance of the ABC algorithm in Fig.~\ref{Fig:non-lin_samp_2_distance}.

\section{Conclusions} 
\label{Sec:Conclusions}
In this paper, we have approached the problem of performing quantum Bayesian inference when the underlying quantum system does not allow for efficient computation of the likelihood function, but the simulation of measurement data still remains tractable thanks to the quantum Monte-Carlo techniques~\cite{Dalibard1992,Carmichael1993,Hegerfeldt1993,Molmer1993,Molmer1996}.  We argue that the reconstruction of the posterior distribution is then still possible by resorting to approximate sampling-based methods.  In particular, upon building a sufficiently large library of measurement data, we demonstrate that likelihood-free approaches---in particular, the Approximate Bayesian Computation (ABC) algorithm~\cite{Turner2012,Sisson2018}---can then be used to reconstruct the posterior. 

In order to show the applicability of our methods, we consider examples of a two-level atom and an optomechanical system, both being driven by classical light and continuously probed by photodetection. The former allows us to study the trade-off between the accuracy and efficiency of the ABC algorithm, thanks to the inference problem being actually solvable analytically. In stark contrast, in the latter case the sampling-based method is often essential to obtain results within a reasonable computation time. Yet, the understanding of the quantum statistics exhibited by the emitted photons~\cite{Clark2022} (bunching/antibunching) is then crucial to choose appropriate statistical tests employed within the ABC algorithm allowing it to distinguish well between different datasets.   

In each case we demonstrate explicitly how the ABC algorithm should be tailored, in particular, we identify the summary statistic of the measured data that is sufficient to characterise the underlying process. Moreover, we study how to choose the thresholds for the acceptance criteria used in rejection sampling, which, if chosen too tight, may require impractically large numbers of samples. For the non-linear optomechanical system we demonstrate that the histogram of waiting times can be used as a summary statistic that is sensitive to photon-bunching effects, in order for the ABC to reconstruct the posterior distribution reasonably well despite minimal calibration. However, let us note that if one (albeit unlikely) requires to reconstruct the posterior in the full parameter range, then ABC can be used to only infer an ``intermediate'' posterior that is already narrowed down to relevant parameter ranges based on the data. As a result, such a narrowed posterior can be used afterwards as a prior by more computationally demanding methods, which can then be already effective thanks to the restricted parameter range.

Apart from the open quantum systems considered here, we emphasise again that the methods presented may be applicable to a wide range of physical systems. More precisely, any system that possesses the \emph{asymmetry} such that it is easy to simulate and sample from its measurement data, but hard to compute the likelihood for, will benefit from the algorithmic approach described in this work. This is the case, for instance, in Hamiltonian learning \cite{Wiebe2014,Wiebe2014pra}, but one may also think of other scenarios in which combining quantum Monte-Carlo techniques~\cite{Dalibard1992,Carmichael1993,Hegerfeldt1993,Molmer1993,Molmer1996} with other methods deems sampling easy while the likelihood computation remains hard. For example, it is so for many-body systems that can be mapped onto spin models~\cite{Manzoni2017}, so that the quantum-jump dynamics can then be efficiently simulated with help of matrix-product-state (MPS) representations~\cite{Lesanovsky2013,Daley2014,Wall2016}.
 
For the above reasons, we believe that our results pave the way for using likelihood-free inference methods across different experimental platforms. A natural next step is to verify their performance in particular quantum estimation tasks, while combining them with other statistical inference techniques such as the Markov chain Monte Carlo methods and the Metropolis-Hastings algorithm, as well as particle filters~\cite{Murphy2012}. These are widely used to avoid the need of considering the full parameter space that, in contrast to our work, may pose a serious challenge when inferring multiple (multidimensional) parameters from the measurement data. We leave this as an interesting line of research to be pursued in the future. \vspace{3mm}

\section*{Acknowledgements}
We thank Stephen Johnson, Klaus M\o{}lmer, M\u{a}d\u{a}lin Gu\c{t}\u{a} and John Calsamiglia for many helpful comments. Our work was supported by the Foundation for Polish Science within the “Quantum Optical Technologies” project carried out within the International Research Agendas programme co-financed by the European Union under the European Regional Development Fund. Project C’MON-QSENS! is supported by the National Science Centre (2019/32/Z/ST2/00026), Poland under QuantERA, which has received funding from the European Union's Horizon 2020 research and innovation programme under grant agreement no 731473. LAC also acknowledges project CZ.02.01.01/00/22\_008/0004649 (QUEENTEC) from the Czech Republic's MEYS, supported by the EU and grant 21-13265X of the Czech Science Foundation. The research was also funded in whole or in part by the National Science Centre, Poland under grant no. 2023/50/E/ST2/00457. For the purpose of Open Access, the authors have applied a CC-BY public copyright licence to any Author Accepted Manuscript (AAM) version arising from this submission.

\bibliographystyle{myapsrev4-2}
\bibliography{ABC_bib}

\appendix
\section{Derivation of the relationship between average waiting time and average total time}~\label{Sec:App_waiting}
In this appendix we derive the expectation value for the total time to emit a fixed number of photons for a system that resets into a common state after emission.  We begin by noting that the total time to emit $N$ photons is the sum of the waiting times between emissions, i.e.~that of Eq.~(\ref{Eq:tN}).  Then, we may write the expectation value of this variable as
\begin{align}
	\langle t_N \rangle = & \int \left(\prod\limits_{i=1}^N \dd \Delta t_i \, w(\Delta t_i) \right) \sum\limits_{i=1}^N \Delta t_i \nonumber \\
	= & N \int \dd \Delta t \, w(\Delta t) \, \Delta t \nonumber \\
	= & N \langle \Delta t \rangle \, .
\end{align}
Here we have utilised the fact that $w(\Delta t_i)$ is the same after each emission.  The result is as one would expect, that the expected time to emit $N$ photons is simply $N$ multiplied by the expected waiting time for one photon.

Similarly, we may calculate the variance of the total time to emit $N$ photons in terms of the single photon waiting time.  Firstly, we find
\begin{align}
	\langle t_N^2 \rangle = & \int \left(\prod\limits_{i=1}^N \dd \Delta t_i \, w(\Delta t_i) \right) \left(\sum\limits_{k = 1}^N \Delta t_k \right)^2 \nonumber \\
	= & \int \left(\prod\limits_{i=1}^N \dd \Delta t_i \, w(\Delta t_i) \right) \sum\limits_{k,l = 1}^N \Delta t_k \, \Delta t_l \nonumber \\
	= & \int \left(\prod\limits_{i=1}^N \dd \Delta t_i \, w(\Delta t_i) \right) \left(\sum\limits_{k=l} \Delta t_k^2 + \sum\limits_{k \neq l} \Delta t_k \Delta t_l \right) \nonumber \\
	= & N \langle (\Delta t)^2 \rangle + 2 \begin{pmatrix} N \\ 2 \end{pmatrix} \langle \Delta t \rangle^2 \nonumber \\
	= & N \langle (\Delta t)^2 \rangle + N (N-1) \langle \Delta t \rangle^2 \, .
\end{align}
Then we calculate the variance to be
\begin{align}
	\mathrm{Var} [t_N] = & \langle t_N^2 \rangle - \langle t_N \rangle^2 \nonumber \\
	= & N \langle (\Delta t)^2 \rangle + N (N-1) (\Delta t)^2 - N^2 \langle \Delta t \rangle^2 \nonumber \\
	= & N \langle (\Delta t)^2 \rangle - N \langle \Delta t \rangle^2 \nonumber \\
	= & N \left( \langle (\Delta t)^2 \rangle - \langle \Delta t \rangle^2 \right) \nonumber \\
	= & N \, \mathrm{Var}[\Delta t] \, .
\end{align}
We have thus shown that the variance of the expected time to emit $N$ photons is related to the variance of the single photon waiting time by a factor of $N$.

\end{document}